# On-demand Mobility-as-a-Service platform assignment games with guaranteed stable outcomes

**Bingqing Liu, Joseph Y. J. Chow[*]**
C2SMARTER University Transportation Center, Department of Civil & Urban Engineering
New York University Tandon School of Engineering, Brooklyn, NY 11021
[*]Corresponding author's email: joseph.chow@nyu.edu

## Abstract

Mobility-as-a-Service (MaaS) systems are two-sided markets, with two mutually exclusive sets of agents, i.e., travelers/users and operators, forming a mobility ecosystem in which multiple operators compete or cooperate to serve customers under a governing platform provider. This study proposes a MaaS platform equilibrium model based on many-to-many assignment games incorporating both fixed-route transit services and mobility-on-demand (MOD) services. The matching problem is formulated as a convex multicommodity flow network design problem under congestion that captures the cost of accessing MOD services. The local stability conditions reflect a generalization of Wardrop's principles that include operators' decisions. Due to the presence of congestion, the problem may result in non-stable designs, and a subsidy mechanism from the platform is proposed to guarantee local stability. A new exact solution algorithm to the matching problem is proposed based on a branch and bound framework with a Frank-Wolfe algorithm integrated with Lagrangian relaxation and subgradient optimization, which guarantees the optimality of the matching problem but not stability. A heuristic which integrates stability conditions and subsidy design is proposed, which reaches either an optimal MaaS platform equilibrium solution with global stability, or a feasible locally stable solution that may require subsidy. For the heuristic, a worst-case bound and condition for obtaining an exact solution are both identified. Two sets of reproducible numerical experiments are conducted. The first, on a toy network, verifies the model and algorithm, and illustrates the differences local and global stability. The second, on an expanded Sioux Falls network with 82 nodes and 748 links, derives generalizable insights about the model for coopetitive interdependencies between operators sharing the platform, handling congestion effects in MOD services, effects of local stability on investment impacts, and illustrating inequities that may arise under heterogeneous populations.

Keywords: many-to-many assignment game, Mobility-as-a-Service, Mobility-on-Demand, Branch and bound, Frank-Wolfe, Lagrange relaxation, subsidy

# 1 Introduction

With the development of information and communication technologies (ICT), new mobility services are emerging rapidly in recent years, such as bikeshare, micromobility, carshare, ride-hail and shared taxis, microtransit, and peer-to-peer ridesharing. Urban passenger transportation is being reshaped from a car-based paradigm to a multimodal one (Shaheen et al., 2020; Chow, 2018). The emergence of new mobility services gave rise to the concept of Mobiltiy-as-a-Service (MaaS) (Djavadian and Chow, 2017; Wong et al., 2020; Pantelidis et al., 2020), which provides mobility services through a joint digital or cyberphysical gateway that enables users to plan, book, and pay for multiple types of mobility services. Recent examples of such platforms include *Moovit*, *Moovel*, *Whim*, *Masabi*, *HaCon*, *Cubic*, *FlexDanmark*, *MovePGH*, and *Cal-ITP*, whose roles are not to operate mobility services but to serve as aggregators of different services (van den Berg et at., 2022; Xi et al., 2022; Zhou et al., 2022). A "platform" refers not to software, but to the broader definition of a type of "two-sided" market where buyers and sellers are simultaneously managed (Rochet and Tirole, 2006), which we adopt in this study.

With more mobility providers entering the market, there's a growing need for managing the operation of mobility providers in MaaS platforms including Mobility-on-Demand (MOD) services that use ICTs. Existence of traditional transit and MOD providers result in MaaS ecosystems with higher levels of flexibility and volatility (Chow, 2018; Djavadian and Chow, 2017; Wong et al., 2020; Pantelidis et al., 2020), potentially generating greater social welfare (Zhou et al., 2022). MaaS platforms are two-sided markets with two mutually exclusive sets of agents, i.e. users and operators, while the platform acts as a regulator. Evaluation of MaaS platforms depends on users' route choices, operators' operation decisions, and platform's pricing and policies. The emergent platform paradigm allows for such analysis.

However, the traditional way of modeling multiplayer markets as generalized Nash equilibria does not fit the goal of the MaaS platform, which would assume a purely noncooperative behavior between operators. Given that the platform sets the fares (either path-based or as subscriptions), it is possible for the platform to achieve more efficient and cost-effective equilibria by fare setting. Rasulkhani and Chow (2019) proposed a game-theoretic platform-based model of MaaS systems with unimodal trips, where each route is operated by one operator. Pantelidis et al. (2020) proposed a many-to-many assignment game method to model MaaS systems with multimodal trips, where each route can be operated by one or more fixed-route operators. The model consists of a matching problem modeled as a multicommodity capacitated fixed-charge network design problem (MCND) (Magnanti and Wong, 1984; Gendron and Larose, 2014), minimizing the sum of travel cost and operating cost (i.e. system cost). The problem's optimal solution is a set of stable outcomes that determines the pricing mechanisms to ensure stability of the assignment.

There are three unresolved problems in the earlier work. First, only fixed-route mobility providers are modeled in the research above, which leaves out analysis of MOD operations as well as multimodal options like fixed route trips with MOD last mile services. MOD providers operate within and between selected service regions with a selected fleet size. Instead of hard capacities, MOD services are characterized by flow-dependent congested wait times. Second, while earlier studies do allow for elastic demand where users choose not to participate in the system, this feature was never explored in depth to understand the consequences. Third, while Sotomayor (1992) has shown that many-to-many assignment games in general exhibit nonempty stable outcome spaces, the proposed problem with MOD services has a more complex structure which makes it unclear whether the nonempty property holds. We show that empty stable outcome spaces are possible, which would lead to an empty stable outcome space issue for the optimal assignment. Pantelidis et al. (2020) did not propose any solution method to solve the matching problem, much less address stability constraints.

We propose new model extensions that substantially expand the applicability of the framework to MaaS analysis and design efforts with MOD services. The matching model from Pantelidis et al. (2020) is extended into a nonlinear mixed integer programming problem that considers trade-offs between congestion in MOD service, capacities in fixed-route transit, and decisions from travelers and both sets of operators. An exact solution algorithm is proposed to solve the matching problem. The algorithm has a branch and bound framework. Each branch is solved through Lagrangian relaxation and subgradient optimization with a Frank-Wolfe algorithm in each iteration. To deal with nonempty outcome spaces, we modify the algorithm into a bounded heuristic to derive solutions with guaranteed stable outcomes. Two sets of numerical experiments are conducted to verify the model, the algorithms, and evaluate their effectiveness in analyzing a range of MaaS design and analysis scenarios. The larger example is based on an expanded Sioux Falls network with 82 nodes and 722 links that includes four fixed-route operators and three MOD operators considering overlapping service regions.

The paper is structured as follows. Section 2 is the literature review. Section 3 introduces the assignment game models. Section 4 introduces the solution algorithms. Section 5 presents two numerical examples and analyses. Section 6 presents the conclusion.

## 2 Literature review

MaaS platforms consist of three sets of decision-makers: travelers, mobility operators (which can fall into fixed-route services or MOD services), and the platform regulator. The travelers decide which multimodal paths they want to take; the operators decide which routes or zones to serve, the amount of line capacity or fleet sizing, and cost transfers with travelers (e.g. fares, transfers between operating costs and user costs such as using virtual stops with increased access time and reduced routing cost); platform regulators can subsidize different operators or travelers and may have control over the design of the market. In the case of public platform regulators, decisions can further extend to the design of the built environment.

There are few studies that consider multiple operator settings, much less on MaaS platforms. Network flow games (e.g. Derks & Tijs, 1985) tend to consider only noncooperative behavior between operators and ignore travelers' route decisions. Generalized Stackelberg games (e.g. Zhou et al., 2005) that have upper level noncooperative games between operators and lower-level user equilibrium model also only consider noncooperative decisions, assume leader-follower role between operators and travelers, and lack the platform/regulator role. Dandl et al. (2021) proposed a tri-level modeling approach considering a single operator with equilibrium determined through simulation. The leader-follower role makes less sense in MaaS if the system behaves more as a two-sided market. Other simulation-based methods also exist (Djavadian and Chow, 2017; Kucharski and Cats, 2022). Zhou et al. (2022) study the specific case of multiple MOD (ride-sourcing) providers competing for travelers which does not consider potential for collaboration in a common platform. Van den berg et al. (2022) propose models for different operating structures of MaaS but they lack network effects. Najmi et al. (2023) model the equilibrium of multimodal markets with multiple providers as a noncooperative game where every link in a road network has a cost of being served by a different mode/operator.

MaaS platforms can be considered as assignment games. Operators are the "sellers", users are the "buyers", and the platform can find a role in the mechanism to determine the cost allocations, depending on their business model. An assignment game (Shapley and Shubik, 1971; Roth and Sotomayer,1992) is a special case of stable matching (Gale and Shapley, 1962), in which utilities are transferable between buyers and sellers. The matching is determined such that no one has behavioral incentive to break from their match. The availability of transferable utilities results in a space of outcomes of cost transfers between parties that would be stable. The game is defined by a matching problem and its corresponding stability conditions. The matching problem determines optimal matches while the corresponding set of stable cost allocation outcomes ensure that the optimal matching is also stable. Shapley and Shubik (1971) showed for the one-to-one assignment game, the matching problem is a linear program and the stable outcome subproblem is its dual, non-empty, and corresponds to the core. Sotomayor (1992) extended the problem to the "multiple partners game", which is a many-to-many assignment game. She showed that the set of stable outcomes for an optimal assignment is nonempty and is a lattice subset of the core (Sotomayor, 1999).

The assignment game has been adapted to model transportation services as many-to-one and many-to-many assignment games. Each operator offers a service route that can match with multiple travelers up to a capacity (Rasulkhani and Chow, 2019) while travelers can transfer between multiple operators. The option for a traveler to match with multiple operators adds a layer of cooperation into the behavioral dynamics of this setting. Pantelidis et al. (2020) proposed a many-to-many assignment game model in which each buyer group is a distinct origin-destination-path, while each operator owns one or more links in a fixed-route service network (i.e. a link represents service from one origin to a destination, not a road segment). In that model, the matching problem is expanded from the assignment model in the one-to-one case (Shapley and Shubik, 1971) into a multicommodity network design (MCND) problem where buyers/travelers decide paths to take while sellers/operators decide which links to operate. Meanwhile, the complementary stable outcome set is expanded to account for effects of fixed costs of operating a link as well as the binding capacity effects on stability. Note that the matching problem in this case finds the optimal matching from a system perspective, e.g. the MaaS platform/the regulator, instead of the optimal matching for one or more operators or users. Fares are assumed to be determined by the platform/regulator as cost transfers that stabilize the optimal matching coalitions. This allowance for coopetitive behavior (where operators may cooperate to serve a multimodal path) is the major difference from the generalized Nash equilibrium modeling framework widely adopted by the literature (Zhou, et al., 2005.; Yang et al., 2000, 2002, 2010; Yang and Yang, 2011, He and Shen, 2015; Ban et al., 2019; Zhang, K. and Nie, Y.M., 2021) and the assignment game proposed by Pantelidis et al. (2020).

The model from Pantelidis et al. (2020) does not recognize MOD operators and their unique characteristics. For example, a MOD service is typically defined as a fleet operating within a service region in which a traveler engages with the system in real time. There can be congestion for users to access the service: the smaller the fleet, the longer it takes for a traveler to access the service. The earliest example models of MOD services involve taxis, using macroscopic cost functions to capture the impedances of matching passengers with drivers (Yang et al., 2000, 2002, 2010; Yang and Yang, 2011) which have been further extended to general ride-hail services (He and Shen, 2015; Zha et al., 2018; Xu et al., 2021a,b; Zhang and Nie, 2021; Correa et al., 2021; Vignon et al., 2023).

Existence of a stable outcome depends on the complicated relationship between trip utility, travel cost, operating cost, and ownership of different parts of the network. If some matching pairs do not provide enough gain to be allocated between users and operators, the users may deviate from the routes selected in optimal matching or just end up unserved. Such issues can be solved by subsidizing the unstable matching pairs. More precisely, the MaaS platform/agencies can intervene by "injecting" subsidies to the unstable matching pairs to increase their potential gain to make sure that no user deviates from the matching. Tafreshian and Masoud (2020) used a minimum subsidy problem to obtain a stable outcome for a match with an empty core in peer-to-peer ridesharing matching games, which shows that subsidization is an effective way of stabilizing matchings.

The most similar prior study is Xi et al. (2022), who proposed a bilevel problem in which the lower-level model is a two-sided market. Like our study, they make use of a branch-and-bound type algorithm and assume linear wait time functions for MOD services. There are several key differences from our work. First, their model does not consider multimodal collaboration between multiple operators to serve a single passenger; in other words, this assumption likens their model to a "transportation problem" compared to our "multicommodity flow problem" with transshipment locations. Second, their model assumes operators and platform regulators that use a specific cost allocation mechanism (NYOP auction) whereas our model is mechanism-agnostic. Third (and most importantly), their model lacks a spatial component: i.e. travelers do not have specific origins and destination; network effects capture only a wait time component that depends on all users without considering link capacities or proximity to other zones and providers. Their model can be an effective tool for high level policy decision support, but limits applicability to network analysis and design.

## 3 Proposed MaaS platform assignment game model

The goal of this study is to build a model that recognizes congested MOD operations so that trade-offs can be analyzed regarding routes served and capacities, service regions and fleet sizes, market entry, elastic user demand for the platform, and subsidies from the platform. The modeling framework takes the original model in Pantelidis et al. (2020) and significantly expands upon that. As the model considers both traveler assignment as well as operator design decisions and platform policies, we call this more generalized model a "MaaS Platform Assignment Game".

Consider a network managed by the MaaS platform initially defined as a directed graph $G(N, A)$ serving a set of traveler groups $S$ traveling to/from centroids $N_Z \subset N$. Travelers in one traveler group $s \in S$ are composed of a homogeneous population of travelers $d_s$, which must have the same origin-destination (OD) pair. Heterogeneous travelers with the same OD pair can be split into different traveler groups to represent different socio-demographic characteristics, such as different income levels and age groups. The MaaS platform has 2 sets of operators: fixed-route operators $Q_F$ and MOD operators $Q_M$.

**Fixed-route transit network**

There exists a set of fixed-route operators $Q_F$ such that each operator $f \in Q_F$ owns one or more service links, $A_f \subseteq A_F$. A node connecting two service links owned by different operators can be expanded using transfer links $l \in A_0$ owned by no operator. $A_{0f}$ denotes the transfer links to the services provided by operator $f \in Q_F$. Each fixed-route operator $f$ decides whether to operate link $l \in A_f$ between nodes $i$ and $j$, and at what service frequency to operate on link $l \in A_f$. To model such decisions, multiple links $l \in A_{ijf}$ ($A_{ijf} \subseteq A_f, A_f \subseteq A_F$) controlled by $y_l \in \{0,1\}$ are created connecting the same two nodes $i$ and $j$ to represent the frequency or line capacity options. Each parallel link $l \in A_{ijf}$ connecting $i,j$ is associated with a capacity $w_l$, an operating cost $c_l$, and a travel cost $t_l$. Note that link travel cost here includes cost of in-vehicle travel time and average wait time which is determined by service frequency. Parallel links with different frequencies have different travel cost and operating costs because increased frequency would reduce average wait time and increase operating cost. At most one of the parallel links can be chosen in each $A_{ijf}$. In practice, additional constraints can be added to fulfill other constraints of capacity selection, for example, ensuring that the same frequency/capacity is selected for service links that belong to the same line.

**MOD network**

MOD operators $Q_M$ operate in regions instead of lines. Service region design can be modeled as choices of whether to provide service in a set of zones and choices of fleet size to deploy in the chosen zones. Alternative service zones of MOD operator $f \in Q_M$ are modeled as nodes using zone centroids $i \in N_{fz}, f \in Q_M$. To represent wait/access cost of MOD services, MOD nodes are created separately from the MOD zone centroids; MOD zone centroids and fixed route links are all connected to origin-destination zone centroids through access/wait links.

MOD services are provided between any pair of zones in the designed service region with a fleet size, which can be represented as complete subgraphs connecting all served zone centroids in the service region. If different fleet size options $h \in H_f$ are considered by an MOD operator $f$ for a service region, different layers of complete subgraphs connecting different layers of MOD nodes are created to represent different fleet size options, each are connected with the MOD zone centroids with MOD access links. An MOD operator $f$ operates with fleet size $h \in H_f$ in a complete subgraph of nodes $N_{fh}$ connected to every other node via links $A_{fh} = N_{fh} \times N_{fh}$. In this case, each MOD node represents the combined choice of a zone and a fleet size as $v_i \in \{0,1\}, i \in N_{fh}$. If a MOD node $i \in N_{fh}$ is chosen, a fixed cost $q_i$ is added to the operating cost of the MOD operator, representing the installation cost of relevant infrastructure. MOD access links $A_{0fh}$ connect zone centroids $N_{fz}$ to MOD nodes $N_{fh}$, representing a wait/access cost given fleet size $h \in H_f$. The MOD access link set of operator $f \in Q_M$ is $A_{0f} = \bigcup_{h \in H_f} A_{0fh}$, where the MOD access link set of all MOD operators is $A_{0M} = \bigcup_{f \in Q_M} A_{0f}$. The MOD link set of MOD operator $f \in Q_M$ is $A_f = \bigcup_{h \in H_f} A_f$ and MOD link set of all MOD operators is $A_M = \bigcup_{f \in Q_M} A_f$. The MOD node set of MOD operator $f \in Q_M$ is $N_f = \bigcup_{h \in H_f} N_{fh}$. The MOD link set of all MOD operators is $N_M = \bigcup_{f \in Q_M} N_f$. A link set that consists of all links in the network is defined as $A = A_F \cup A_0 \cup A_M \cup A_{0M} \cup A_D$.

Fig. 1 shows how to develop the network combining different alternative fixed-route frequency and MOD service fleet designs. If detailed demand with OD pairs of coordinates are modeled, street networks can be used to connect origins, destinations, fixed-route transit stations, and MOD service zone centroids, although computational trade-offs will need to be considered. Fig. 1 shows an example of the latter.

**Matching optimization**

From a system perspective of the MaaS platform, we want to find the set of matches that minimizes system cost, which is the sum of the total cost of travelers in the platform and total operating cost of operators. Fares are transferred costs from travelers to operators, which cancel out from the system perspective (similar to how toll revenues are assumed to feed back into the social surplus in congestion pricing). We name such a set of matches as the *system-optimal matching*, which involves finding the link flows $x_{sl}$ of link $l \in A$ per user group $s \in S$ such that the sum of travelers' total costs and operators' total costs is minimized. Such an objective maximizes operator's total profit and travelers' total social welfare (per Shapley and Shubik, 1971) implicitly by minimizing total costs, since fares are transferred costs from travelers to operators which cancels out from the system perspective (see Rasulkhani and Chow, 2019). Travelers of user group $s \in S$ are assumed to gain utility $U_s$ upon completing their trip (typically dependent on the trip purpose). Ma et al. (2021) provide an example of calibrating these utilities from existing travel modes. Note that stability is not ensured by finding the system-optimal matching, but by the stability conditions in Section 3.2 while finding a non-empty set of fares, which are equivalent to equilibrium conditions (with pricing) as shown in Section 3.3.3.

To allow for users to opt out of the MaaS platform, the OD pair of each traveler group $s \in S$ is connected by an uncapacitated dummy link $l \in A_D$, associated with a travel disutility $t_s \leq U_s$, 0 operating cost, and infinite capacity. Travelers using these links are assumed to be not participating in the platform, instead using external modes, a competing platform, or not traveling. Elastic user demand for the platform is captured through these dummy links. If a choice model was estimated as a logit model and the respective utility functions normalized to the common currency, then the travel disutility $t_s$ can be calibrated as the logsum of these external modes' utility functions to be integrated as a deterministic alternative.

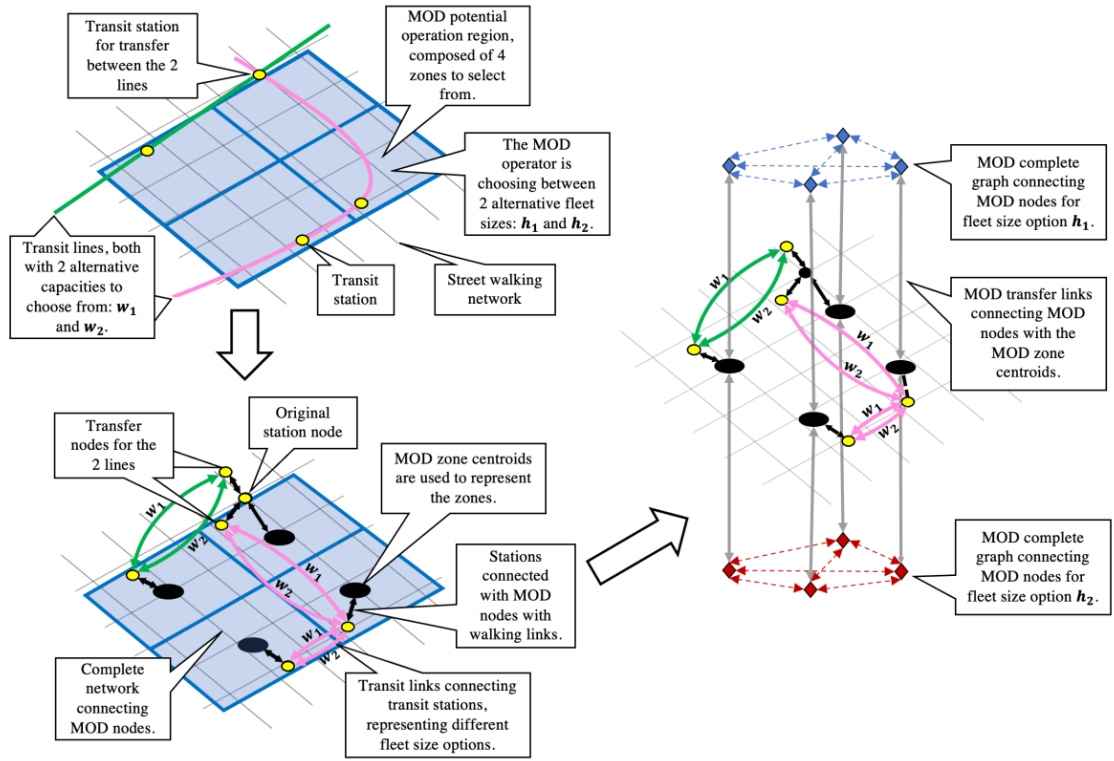

Figure 1. Network structure illustration.

For MOD access links, disutilities of wait/access time are modeled differently from the fixed-route operations. Similar to the literature on matching functions for taxi and ride-hailing services, $\tau_l$ is used for modeling traveler disutility as shown in Definition 1.

**Definition 1**. *The matching friction between travelers and MOD services at MOD service zones is modeled as the travel disutilities of MOD access links $l \in A_{0M}$, representing the steady-state average access/wait times at the MOD service zones. A macroscopic disutility function $\tau_l(\sum_s x_{sl}; h)$ is defined by Eq.(1).*

$$\tau_l\left(\sum_s x_{sl}; h\right) = a_1 \left(\sum_s x_{sl}\right)^{b_1} h^{b_2} \qquad (1)$$

For MOD access links, travel disutilities are nonlinear, exhibiting congestion effects due to queueing for MOD service but no capacity effects. For MOD operators, their operating costs include supporting travelers to traverse MOD links and infrastructure cost of MOD nodes. The infrastructure cost of a MOD node represents the cost of setting up a zone for service for a given fleet size $h$ (e.g. bike dock installation for bikesharing services), which is defined as a constant $q_i$ for MOD node $i \in A_M$.

Operating costs of links are also defined similarly from the literature as shown in Definition 2.

**Definition 2.** *Unit operating cost of MOD link $l \in A_M$ between two served zones is defined as $m_l$, which is the operating cost of transporting 1 unit of demand between two served zones. It is calibrated by the fleet size of the operator and other possible factors (e.g. link travel cost) as shown in Eq. (2), where $a_2 > 0$ and $b_3 > 0$. The parameters imply that as fleet size increases, the unit operating cost increases.*

$$m_l(h) = a_2 h^{b_3} \qquad (2)$$

Since fleet size $h$ is a constant for a MOD link $l \in A_M$ given the network structure, the unit operating cost for traversing two zones is a calibrated constant that varies for different fleet sizes.

Since the assignment game assumes a transferable-utility game, all utility-related parameters use a common currency exchangeable between travelers and operators (e.g. dollars). Due to the complexity of the model settings, a list of assumptions is listed below. A list of notations is provided as Table 1.

List of assumptions:
- The proposed method models the MaaS platform from a planning perspective.
- The MaaS platform sets fares for the services it owns. Operators within the MaaS platform give up their right of fare setting.
- Travel costs of fixed-route transit links are fixed, including the cost of in-vehicle travel time and average wait time determined by service frequency. We do not consider the variations of travel costs of these links caused by traffic status.
- Travel cost of a transfer link connecting fixed-route services is the cost of walking.
- Traveler congestion in MOD services caused by limited MOD fleet size is captured by a MOD access link cost function.
- MOD in-vehicle travel costs between MOD zones are considered fixed. We consider the traffic congestion on road network caused by MOD operation negligible relative to the background traffic.
- MOD in-vehicle travel costs are captured by travel costs of MOD links, which represent the travel cost of the shortest path or the generally used path connecting two MOD zones.
- All travel costs are non-monetary costs, not including fares.
- Operating cost of fixed-route transit links is fixed, associated with capacities of the links.

- Operating cost of MOD includes the zone opening cost corresponding to the installation costs of required infrastructure (e.g. bike dock), and costs of operating the fleet on MOD links.

Table 1. List of Notations

| | |
|---|---|
| **System notations** | |
| $N$ | Set of nodes |
| $A$ | Set of links |
| $S$ | Set of traveler groups |
| $A_f$ | Set of links owned by operator $f$ |
| $A_0$ | Set of transfer links owned by no operator |
| $A_{0f}$ | Set of transfer links accessing the service provided by operator $f$ |
| $A_D$ | Set of uncapacitated dummy links connecting each OD pair, representing users that opt out of the MaaS platform |
| $U_s$ | Utility a traveler in traveler group $s$ gains upon completing the trip |
| $d_s$ | Travel demand of traveler group $s$ |
| $R_s$ | Set of all paths connecting OD pair of traveler group $s$ |
| $R_s^*$ | Set of optimal paths solved from $L_1$ connecting OD pair of traveler group $s$ |
| $A_r$ | Set of links that compose path $r$ |
| $A_i^+$ | Set of outbound links of node $i$ |
| $A_i^-$ | Set of inbound links of node $i$ |
| **Fixed-route transit notations** | |
| $Q_F$ | set of fixed-route operators |
| $A_F$ | Set of links owned by all fixed-route operators |
| $A_{ijf}$ | Set of parallel fixed-transit links connecting node $i$ and $j$ to represent the frequency or line capacity options |
| $w_l$ | Link capacity of fixed-route link $l$ |
| $c_l$ | Operation cost of fixed-route link $l$ |
| $t_l$ | Travel cost of fixed-route link $l$, including cost of in-vehicle travel time and average wait time determined by service frequency |
| $\mu_l$ | Lagrange multiplier of fixed-route link $l$ |
| **MOD notations** | |
| $Q_M$ | Set of MOD operators |
| $N_{fz}$ | Set of centroids of the service zones owned by MOD operator $f$, connected with fixed-route transit links by links in $A_0$ |
| $H_f$ | Set of fleet size options considered by MOD operator $f$ |
| $N_{fh}$ | Set of nodes connected with each other (complete graph) representing the option of MOD operator $f$ serving with fleet size $h$ |
| $A_{fh}$ | $N_{fh} \times N_{fh}$, set of links constructing complete graph between nodes in $N_{fh}$, representing the option of MOD operator $f$ serving with fleet size $h$ |
| $A_{0fh}$ | Set of MOD access links connecting zone centroids $N_{fz}$ to MOD nodes $N_{fh}$, representing a wait/access cost given fleet size $h$ |
| $A_{0M}$ | $\bigcup_{f \in Q_M} A_{0f}$, set of MOD access links |
| $A_M$ | $\bigcup_{f \in Q_M} A_f$, set of MOD links |
| $N_f$ | $\bigcup_{h \in H_f} N_{fh}$, MOD node set of MOD operator $f$ |
| $N_M$ | $\bigcup_{f \in Q_M} N_f$, MOD link set of all MOD operators |
| $\tau_l$ | Travel disutility of MOD access links $l \in A_{0M}$, representing the steady-state average access/wait times at the MOD service zones, function of link flow $\sum_s x_{sl}$ and fleet size $h$ |

| $m_l$ | Unit operating cost of MOD link $l$, function of fleet size $h$ |
| $q_i$ | Node opening cost of MOD node $i \in N_{fh}$ |
| **Decision variables** | |
| $x_{sl}$ | Continuous decision variable of link flow of link $l$ per traveler group $s$ |
| $y_l$ | Binary decision variable of whether a fixed-route link is chosen to be operated |
| $v_i$ | Binary decision variable of whether a MOD node $i$ representing a MOD service zone and a fleet size $h$ is chosen |
| $p_l$ | Continuous decision variable of link fare |
| $u_s$ | Continuous decision variable of traveler's payoff of traveler group $s$ |

### 3.1 Matching model considering fixed-route and MOD operators

The assignment game under these changes has the following nonlinear integer programming problem $L_1$ in Eq. (3) as the new system-optimal matching problem.

*Matching problem $L_1$*:

$$\min Z_{L_1} = \sum_{s \in S} \sum_{l \in A_0 \cup A_F \cup A_M \cup A_D} t_l x_{sl} + \sum_{s \in S} \sum_{l \in A_{0M}} \tau_l \left( \sum_{s \in S} x_{sl} ; h \right) x_{sl} + \sum_{l \in A_F} c_l y_l + \sum_{s \in S} \sum_{l \in A_M} m_l x_{sl} \quad (3a)$$
$$+ \sum_{i \in N_M} q_i v_i$$

subject to
$$\sum_{l \in A_i^+} x_{sl} - \sum_{l \in A_i^-} x_{sl} = \begin{cases} d_s, i \in O_s, \\ 0, \text{ otherwise}, \\ -d_s, i \in D_s, \end{cases} \forall i \in N, \forall s \in S \quad (3b)$$

$$\sum_{s \in S} x_{sl} \leq w_l y_l, \forall l \in A_F \quad (\mu_l) \quad (3c)$$

$$\sum_{s \in S} x_{sl} \leq \sum_{s \in S} d_s v_i, \quad \forall i \in N_M, \forall l \in A_i^+ \quad (3d)$$

$$\sum_{l \in A_{ijf}} y_l \leq 1, \forall A_{ijf} \subseteq A_F \quad (3e)$$

$$v_i + v_j \leq 1, \forall i \in N_{fh}, \forall j \in N_{fh'}, \forall (h, h') \in H_f \times H_f, h \neq h', \forall f \in Q_M \quad (3f)$$
$$x_{sl} \geq 0, \quad \forall l \in A, \forall s \in S \quad (3g)$$
$$y_l \in \{0,1\}, \quad \forall l \in A_F \quad (3h)$$
$$v_i \in \{0,1\}, \quad \forall i \in N_M \quad (3i)$$

The objective is the system cost, which includes travelers' total travel cost and operators' total operating cost. Eq. (3b) are the flow conservation constraints, where the set of outbound links of node $i$ is defined as $A_i^+$, and the set of inbound links of node $i$ is defined as $A_i^-$. Eq. (3c) are the capacity constraints with Lagrange multiplier $\mu_l$. Eq. (3d) are the MOD node controlling constraints, which ensure that flows can only exist between two MOD nodes if both nodes are opened. Only when $v_i = 1$, can the flows on the outbound links of MOD node $i$ be positive. Eq. (3e) ensures that only one service frequency/capacity is chosen for each fixed-route link $(i, j)$ by operator $f$. Eq. (3f) ensures that only one fleet size is chosen by each MOD operator. When $v_i = 1$, then $v_j = 0$ for all other nodes $j$ owned by the same operator as $i$ but associated with different fleet sizes. Eqs. (3g) – (3i) are non-negativity and binary constraints. The model is a nonlinear integer program; when travel costs are constant, it simplifies to an NP-hard MCND problem (Magnanti and Wong, 1984). This belongs to the class of convex multicommodity network design problems (Crainic and Rousseau, 1986; Paraskevopoulos et al., 2016), although those tend to consider a system optimal formulation and include both congestion and capacity effects on the same links whereas Eq. (3) does not.

We seek to minimize the system cost for two reasons. First, modeling MaaS platform's perspective requires considering both users and operators as a system. However, fare is a payoff/cost transferred from the users to the operators which cancels out from a system perspective. That is the reason why we do not seek to maximize any operators' profit when solving for the matching. The second reason is that from a system perspective, minimizing total costs implicitly maximizes the potential profit of operators by maximizing total payoff of users and operators. Per Rasulkhani and Chow (2019), An optimal solution (x, y, v) to Eq. (3) ensures maximal space for cost allocation, which further ensures maximal potential revenue. Based on the optimal matching solved from Eq. (3), fares can be set to maximize operators' profit or users' payoff. In Section 3.2, we will discuss fare setting and cost allocation in detail.

**3.2 Corresponding stability conditions**

Note that the stability in game theory, which is adopted in this paper, is not the same as the stability in in optimization or control. The latter means that a small perturbation (often in data) would not dramatically change the model solution. In our case, stability of a match means that there is no incentive for any player to change choices.

The stable outcome space corresponding to $L_1$ is shown in Eqs. (4a) – (4f). In this study we focus on link-additive pricing policy. We denote $p_l$ as the fare of link $l$ and $u_s$ as the payoff of a user on OD pair $s \in S$. The variables $x_{ls}^*, y_l^*, v_i^*$ and $\mu_l^*$ represent the optimal flow on link $l \in A$ on OD pair $s \in S$, optimal operation choice of fixed-route link $l \in A_F$, optimal operation choice of MOD node $l \in N_M$, and Lagrange multiplier of fixed-route link $l \in A_F$ from the matching problem $L_1$. $R_s^*$ is the set of optimal paths for user $s \in S$ corresponding to the solution of Eq. (3). $R_s$ is the set of all paths connecting OD pair $s \in S$. $A_r$ denotes the set of links that compose matched paths $r \in R_s^*, s \in S$. $A_{r\prime}$ denotes the set of links that compose an unmatched path $r' \in R_s \backslash R_s^*, s \in S$. Eqs. (4a) – (4b) represent minimum operator costs (individual rationality for each operator). Eq. (4c) ensures utility conservation. Eq. (4d) ensures that an outcome is locally stable, i.e. there is no incentive for a single traveler to switch to another path.

$$\sum_{l \in A_f} p_l \sum_{s \in S} x_{sl}^* \geq \sum_{l \in A_f} c_l^* y_l^*, \forall f \in Q_F \tag{4a}$$

$$\sum_{l \in A_{0f}} p_l \sum_{s \in S} x_{ls}^* \geq \sum_{l \in A_f} \sum_{s \in S} m_l x_{sl}^* + \sum_{i \in N_f} q_i v_i^*, \forall f \in Q_M \tag{4b}$$

$$u_s + \sum_{l \in A_r} p_l = U_s - \left( \sum_{l \in A_r \cap (A_0 \cap A_F \cup A_D \cup A_M)} t_l + \sum_{l \in A_r \cap A_{0M}} \tau_l \left( \sum_{s \in S} x_{ls}^*; h \right) \right), \forall r \in R_s^*, s \in S \tag{4c}$$

$$u_s \geq U_s - \left( \sum_{l \in A_{r\prime} \cap (A_0 \cup A_F \cup A_D \cup A_M)} (t_l + \mu_l^* + c_l(1 - y_l^*)) + \sum_{l \in A_{r\prime} \cap A_{0M}} \tau_l \left( \sum_{s \in S} x_{ls}^* + 1; h \right) \right.$$
$$\left. + \sum_{l \in A_{r\prime} \cap A_M} m_l + \sum_{i \in N_{r\prime} \cap N_M} q_i(1 - v_i^*) + \sum_{l \in A_{r\prime}} p_l \right), \forall r' \in R_s \backslash R_s^*, r \in R_s^*, s \in S \tag{4d}$$

$$u_s \geq 0, \forall s \in S \tag{4e}$$

$$p_l \geq 0, \quad \forall l \in A_r, r \in R_s^*, s \in S \tag{4f}$$

The passenger path flow stability condition shown in Eq. (4d) is proven to correspond to the fixed-route operator market assignment in Pantelidis et al. (2020) under a more complex non-additive fare policy. With added MOD operators, the stability condition is trivially extended but stated formally here for sake of completeness.

**Proposition 1**. Solution stability. *An optimal solution $(x, y, v)$ to Eq. (3) is locally stable if there exists one or more solutions $(u, p)$ that satisfy Eqs. (4a) – (4f), where $u$ is the vector of utility transferred to travelers and $p$ is the utility vector normalized as the fare payment from travelers to operators*.

*Proof*. The operator cost constraints (Eq. (4a)-(4b)) and utility conservation constraints (Eq. (4c)) are feasibility conditions. Eq. (4d) ensures the local stability of the matched paths. Pantelidis et al. (2020) shows the case for only fixed-route operators where traveler paths may switch to alternative paths. Here we follow the same logic considering two possible conditions of switching one user (local stability consideration) from the matched path to the alternative unmatched path:

Condition 1: The unmatched path is fully operated, including fixed-route links and MOD nodes and links. In this case, similar to Pantelidis et al. (2020), if a capacitated fixed-route link $l$ on the unmatched path is already at capacity, switching one user in requires pushing someone else off at the capacity price $\mu_l^*$.

Condition 2: The unmatched path is not fully operated. The switched user needs to pay for the following costs:
- Fares of the operated links on the unmatched path ($p_l$);
- Operating cost of the link for unoperated fixed-route link $l$ on the unmatched path ($c_l(1 - y_l^*)$);
- Opening cost of the unopened MOD nodes on the unmatched path ($q_i(1 - v_i^*)$);
- Operating cost of unoperated MOD link for one more user ($m_l$);
- Travel cost of fixed-route links, MOD links, and transfer links ($t_l$), including the additional cost of pushing someone else off from the links at capacity ($\mu_l^*$);
- Cost of MOD access links, which is changed to $\tau_l(\sum_{s \in S} x_{ls}^* + 1; h)$ due to the switch.

The above total costs lead to the cost allocation shown in the right-hand-side of Eq. (4d). To ensure stability, the user's payoff $u_s$ solved from the cost allocation of matching path (Eq. (4c)) is much be higher than or equal to the user's payoff of unmatching paths, leading to the inequality of Eq. (4d). ∎

Like Shapley and Shubik (1971) and Pantelidis et al. (2020), the proposed assignment game model is mechanism-agnostic; it does not output a specific cost allocation but instead outputs the set of all stable outcomes corresponding to the optimal matching by identifying two extreme vertices: buyer- and seller-optimal solutions. This is not to be confused with "user equilibrium" and "social optimum" in the transportation assignment literature; in essence every outcome is a "system optimum" with different cost transfers that ensure user equilibrium conditions are met (similar to marginal cost pricing) (see Section 3.3). The vertices can be identified by solving $L_2$ with either Eq. (5a) (buyer-optimal) or (5b) (seller-optimal), where all solutions in between these vertices are also stable.

*Stable outcome problem $L_2$*:
$$\max \sum_{s \in S} u_s \text{ (Buyer optimal)} \quad (5a)$$
or
$$\max \sum_{f \in F} \sum_{l \in A_{0f}} p_l \sum_{s \in S} x_{ls}^* \text{ (Seller optimal)} \quad (5b)$$
Subject to Eqs. (4).

Note that the cost allocation constraints are path-dependent, while the flows solved from $L_1$ are link-based. This will be dealt with in the solution algorithm discussed in Section 4. In addition, not all links are owned by operators in the platform; these include transfer links, access/egress links to stations, and the dummy links connecting OD pairs representing options alternative to the platform.

### 3.3 Model properties and discussion
*3.3.1. Instability for an optimal solution to $L_1$*

While Sotomayor (1992) proved that basic many-to-many assignment games have non-empty stable outcome sets, the proposed MaaS assignment game features network effects that complicate the problem.

With selfish travelers under congestion and the impact on operators, it is possible that the optimal solution to $L_1$ may not satisfy one or more of the constraints in Eq. (4) (Lemma 1).

**Lemma 1.** Empty outcome space. *The assignment game in $L_1$ (Eqs. (3)) is not guaranteed to be stable, i.e. have a non-empty outcome set from Eqs. (4a) – (4f).*

*Proof.* Denote the matched but unstable path as $r$, and the path that users on $r$ have incentives to switch to as $r'$. The two paths connect the same OD pair. Since $r$ is chosen over $r'$ by the matching problem $L_1$, we have Eq. (6). It means that the total cost of the flow on $r$ is less than the cost of allocating the flow to $r'$ (the part of $r$ shared with other matched paths still operating, and the part of $r'$ that is not operating starts operating). Flow on path $r$ is denoted as $x_r$. Set of links on paths $r$ and $r'$ are denoted as $A_r$ and $A_{r'}$, respectively. The set of operating links on $r$ that serves not only users on $r$ but also users on other matched paths is denoted as $A_{r,shared}$. Set of links not operated on the unmatched path $r'$ is denoted as $A_{r',no}$.

$$x_r \sum_{l \in A_r} t_l + \sum_{l \in A_r} c_l < x_r \sum_{l \in A_{r'}} t_l + \sum_{l \in A_{r,shared}} c_l + \sum_{l \in A_{r',no}} c_l \quad (6)$$

The condition of $r$ being unstable with respect to $r'$ is shown as Eq. (7), which is derived by solving $u_s$ from Eq. (4b) and plugging in the solved $u_s$ into Eq. (4c).

$$\sum_{l \in A_r} t_l + \sum_{l \in A_r} p_l \geq \sum_{l \in A_{r'}} (t_l + \mu_l^*) + \sum_{l \in A_{r'}} p_l + \sum_{l \in A_{r',no}} c_l \quad (7)$$

Multiplying Eq. (7) by $x_r$ and combining with Eq. (6), we have Eq. (8), which is the condition of a matched path $r$ being unstable.

$$x_r \sum_{l \in A_r} p_l - \sum_{l \in A_r} c_l > x_r \sum_{l \in A_{r'}} \mu_l^* + x_r \sum_{l \in A_{r'}} p_l - \sum_{l \in A_{r,shared}} c_l + (x_r - 1) \sum_{l \in A_{r',no}} c_l \quad (8)$$

∎

Lemma 1 shows that $L_1$ on its own may not be stable. As such, we define a more constrained version of $L_1$ that includes a local stability guarantee, named as $L_{1C}$. The optimal solution to $L_{1C}$ is a market equilibrium for the platform design that is guaranteed to be stable in Definition 3.

**Definition 3**. *Let $L_{1C}$ be the assignment game defined by $L_1$ plus the added constraint of guaranteed local stability. The optimal solution to $L_{1C}$ is a MaaS platform equilibrium, where $Z_{L_1}^* \leq Z_{L_{1C}}^*$. No user can unilaterally switch path without being worse off, and no operator can unilaterally change their decision without being worse off for any outcome in the non-empty stable outcome set, and cooperative behavior between operators is possible in serving a user path together.*

An example is shown in Fig. 2. There are 2 OD pairs in the network, 1 to 3 and 1 to 2, both with 100 units of demand. Link travel costs and operating costs are labeled in Fig. 2. Links (1,3) and (2,3) are walking links without operating cost and not owned by any operator. Dummy links for the two OD pairs are drawn in yellow with dashed lines with travel costs labeled. Utility $U_s$ of the 2 OD travelers are both 25. All links have infinite capacity. The optimal matched flows solved from $L_1$ are 200 units on link (1,2) and 100 units on link (2,3). However, the matched path for OD1, which is [1,2,3], is an unstable path because of the subsequent cost transfer. The lowest fare that can be set for link (1,2) is 2.4, so the total cost for one user on path [1,2,3] is $12 + 6 + 2.4 = 20.4$, which is greater than the cost of walking path [1,3]. This means that the users on path [1,2,3] have incentives of switching to walking path [1,3]. The condition of Eq. (8) holds in this case (left-hand side = -240, right-hand side = -480).

When instability happens, flow on the unstable path have incentive to switch to unmatched paths. In the example shown in Fig. 2, after the first user switches to walking path [1,3], the fare of link (1,2) will increase, which leads to more users switching to walking path [1,3]. The system stabilizes when all the 100 units of flow on path [1,2,3] have switched to walking path [1,3], which is the MaaS platform equilibrium. However, the equilibrium objective increases after the switch (before: $Z_{L_1}^* = 3{,}480$; after: $Z_{L_{1C}}^* = 3{,}680$).

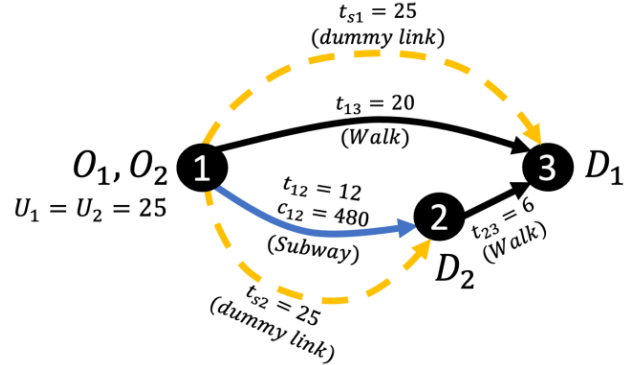

Figure 2. Example of instability.

The stability condition that we consider is local stability, i.e. there is no incentive for one user to switch off his/her current path. It is possible that a locally stable outcome may not be globally stable, i.e. there could be a pair of users that may swap that would lead to a better outcome. This is similar to the difference between a "user equilibrium" and a "user optimum" in Smith (1984), where we align the local stability constraints in Eq. (4) with "user equilibrium", but seek the "user optimum" by optimizing $L_{1C}$ (Proposition 2).

**Proposition 2**. Local and global stability. *A feasible solution to $L_{1C}$ may be locally stable without being optimal (local stability is a necessary but not sufficient condition); the optimal solution to $L_{1C}$ is globally stable.*

*Proof*. As proven by Proposition 1, Eqs. (4a) – (4f) ensures local stability, i.e. there is no incentive for a single traveler to switch to another path.

Given only links with fixed link costs (transfer links and fixed-route links), the cost per user on unmatched path $r'$ when $n$ users switch to unmatched path $r'$ is $\sum_{l \in A_{r'}}(t_l + \mu_l^* + c_l(1 - y_l^*)) + \frac{\sum_{l \in A_{r'}} c_l(1 - y_l^*)}{n}$, which monotonically decreases while $n$ increases, indicating that more users will switch to path $r'$ if the path $r'$ is worthwhile for the first user to switch to. In this case, local stability ensures global stability.

Given links with congestion effects (MOD access links), the cost per user on unmatched path $r'$ when $n$ users switch to unmatched path $r'$ is $U_s - \left(\sum_{l \in A_{r'} \cap (A_0 \cup A_F \cup A_D \cup A_M)}(t_l + \mu_l^*) + \sum_{l \in A_{r'} \cap A_{0M}} \tau_l(\sum_{s \in S} x_{ls}^* + 1; h) + \sum_{l \in A_{r'} \cap A_M} m_l + \frac{\sum_{i \in N_{r'} \cap N_M} q_i(1 - v_i^*) + \sum_{l \in A_{r'} \cap (A_0 \cup A_F \cup A_D \cup A_M)} c_l(1 - y_l^*)}{n}\right)$, in which the last term decrease while $n$ increases, while the second term increases while $n$ increases. In this latter case, it is possible that the lowest payoff per user switched appears when more than one user switches to the unmatched path $r'$, which means that local stability does not ensure global stability when there are congestion effects. Among all the locally stable solutions, the solution with the lowest objective value ($Z_{L_{1C}}^*$) is the MaaS platform equilibrium, which is globally stable, since the lowest objective value in Eq. (3) indicates no incentives of switching in the platform. ∎

### 3.3.2. Unique and non-unique variables
As shown in Sotomayor (1999), the many-to-many assignment game has a unique lattice structure for its stable outcome set and there exists a unique optimal stable payoff for each side of the market (buyer- and seller-optimal vertices). From the network properties, however, Pantelidis et al. (2020) also showed that certain variables are unique even if the path flows are not unique.

**Proposition 3**. Solution uniqueness. *A MaaS platform equilibrium solution under a link-additive and separable pricing scheme is unique in link flows, passenger ridership per operator, and the sum of total consumer surplus and total operator revenue.*

*Proof*. Link flows are known to be unique for multicommodity assignment models under congestion (Sheffi, 1985). As mentioned, we consider two alternative pricing schemes: a link-based price and a non-additive price that is paid once to an operator regardless of how many legs of the trip belong to that operator. The latter scheme was studied in Pantelidis et al. (2020). Here we study the former, which is formulated as the conditions in $L_2$.

Since flow of link $l$ is unique ($\sum_{s \in S} x_{ls}^*$), system travel cost is unique ($\sum_{s \in S} \sum_{l \in A_0 \cup A_F \cup A_M \cup A_D} t_l x_{ls}^* + \sum_{l \in A_{0M}} \tau_l(\sum_{s \in S} x_{ls}^*; h) \sum_{s \in S} x_{ls}^*$). System total trip utility is unique ($\sum_{s \in S} d_s U_s$). Since the sum of system travel cost, system total consumer surplus, and total operator revenue is equal to system total trip utility, the sum of system total consumer surplus and total operator revenue is unique despite the non-uniqueness of path flows. ∎

*3.3.3. Generalization of Wardrop's principles*

The assignment game stability conditions shown in Eq. (4) represent a generalized form of the noncooperative user equilibrium route choice behavior proposed by Wardrop (1952) with added operator considerations and cooperation for operators in serving a user path together.

**Corollary 1**. Generalization of Wardrop's user equilibrium. *When $Q = \{\}$, i.e. there is no operator agent and the platform becomes a one-sided market, the stable outcome space defined by the constraints in Eq. (4) is determined only by $u_s$ ($p_l$'s drop out), which leads to a generalization of the Wardrop's user equilibrium conditions (1952) with known $U_s$ and link capacities.*

*Proof*. Without operators, Eqs. (4a) – (4b) drop out. Removing the operator terms in the stability constraint in Eqs. (4c) – (4d) leads to Eqs. (4c') – (4d').

$$u_s = U_s - \left( \sum_{l \in A_r \cap (A_0 \cap A_F \cup A_D \cup A_M)} t_l + \sum_{l \in A_r \cap A_{0M}} \tau_l \left( \sum_{s \in S} x_{ls}^*; h_l \right) \right), \forall r \in R_s^*, s \in S \quad (4c')$$

$$u_s \geq U_s - \left( \sum_{l \in A_{r'} \cap (A_0 \cup A_F \cup A_D \cup A_M)} (t_l + \mu_l^*) + \sum_{l \in A_{r'} \cap A_{0M}} \tau_l \left( \sum_{s \in S} x_{ls}^* + 1; h_l \right) \right), \forall r' \in R_s \backslash R_s^*, r \in R_s^*, s \in S \quad (4d')$$

If we combine Eqs. (4c') and (4d') and assume that all links are uncapacitated, we have Eq. (9), which is in fact the KKT conditions corresponding to the user equilibrium as shown in Beckmann et al. (1956). The left-hand side is the travel cost of the matched path $r$. The right-hand side is the cost of switching to an unmatched path $r'$, which is travel cost of the unmatched paths.

$$\sum_{l \in A_r \cap (A_0 \cap A_F \cup A_D \cup A_M)} t_l + \sum_{l \in A_r \cap A_{0M}} \tau_l \left( \sum_{s \in S} x_{ls}^*; h \right)$$

$$\leq \sum_{l \in A_{r'} \cap (A_0 \cup A_F \cup A_D \cup A_M)} t_l + \sum_{l \in A_{r'} \cap A_{0M}} \tau_l \left( \sum_{s \in S} x_{ls}^* + 1; h \right), \forall r \in R_s^*, s \in S, \forall r' \in R_s \backslash R_s^*, r \in R_s^*, s \in S \quad (9)$$

### 3.4 Subsidy for system stabilization

The MaaS platform assignment game so far does not involve any decisions from the platform. As solving $L_1$ may not obtain a solution to $L_{1C}$, we consider subsidy decisions for the platform to guarantee local stability. We introduce the role of subsidy in stabilizing an unstable assignment as a platform design strategy first considered by Tafreshian and Masoud (2020) for ridesharing markets. When unstable paths are subsidized sufficiently to enlarge the payoff, they can still be stabilized, and the result may even outperform the solution to $L_{1C}$. In MaaS platforms, users pay one time for a bundle of mobility services which compose a path, so it might make sense for subsidies per user and per path. Consider a path-based subsidy $a_r$ directed to each user on path $r \in R_s^*, s \in S$, such that the travel disutility is decreased by that amount per user in Definition 4.

**Definition 4**. *$L_{1S}$ is a variant of the $L_1$ assignment game with the added decision variable for the platform decision-maker to provide path-based subsidies $a_r$ for path $r \in R_s^*, s \in S$ such that the equilibrium objective is minimized, i.e. in Eq. (10), where $x_r$ is the flow on path $r \in R_s^*, s \in S$.*

$$\min Z_{L_{1S}} = Z_{L_1} + \sum_{s \in S} \sum_{r \in R_s^*} a_r x_r \qquad (10)$$

*The optimal solution to $L_{1S}$ is called a subsidized MaaS platform equilibrium, where $Z_{L_1}^* \leq Z_{L_{1S}}^*$.*

In the example of Fig. 2, instability of path [1,2,3] happens due to larger cost per user ($\sum_{A_r} t_{ij} + \sum_{A_r} p_{ij}$) than the walking path [1,3]. Obviously, such instability can be fixed by injecting subsidies $a_{[1,2,3]}$ to lower the cost of the unstable path [1,2,3]. In this example, a subsidy of $a_{[1,2,3]} = 0.4$ is needed for each user on path [1,2,3] to make sure that they don't switch to walking path [1,3]. Without any intervention, the flow on [1,2,3] will switch to the walking path [1,3], leading to $Z_{L_{1C}}^* = 3,680$. With the subsidy included, the subsidized equilibrium objective is $Z_{L_{1S}}^* = 3,480 + 40 = 3,520$, which is still lower than the equilibrium objective of 3,680. This means that subsidizing the matchings solved from $L_1$ is worthwhile in this case. However, if the cost of link (1,3) is decreased from 20 to 18.5, the flows solved from $L_1$ stays the same, while subsidy needed to stabilize path [1,2,3] becomes $a_{[1,2,3]} = 1.4$ per user. The subsidized equilibrium objective becomes $Z_{L_{1S}}^* = 3,620$. In this case, the equilibrium objective is $Z_{L_{1C}}^* = 3,580$, which means that it is not worthwhile to inject subsidies.

$L_{1S}$ is a highly complex problem since the objective (10) and constraints are nonconvex. We can equivalently decompose the problem into finding a solution to $L_1$ and solving for $a_r^*$ in $L_3$ given the variables obtained from $L_1$ if they are unstable, i.e. Definition 5.

**Definition 5**. *$L_3$ finds the optimum subsidies $a_r$ needed for a given solution $(x^*, y^*, v^*)$ with an empty outcome set such that it is no longer empty. Then $Z_{L_{1S}}^* = \min(Z_{L_1}(x^*, y^*, v^*) + Z_{L_3|(x^*,y^*,v^*)})$.*

*Minimum Subsidy Problem $L_3$:*

$$\min_{a_r, u_s, p_l} Z_{L_3|(x^*,y^*,v^*)} = \sum_{s \in S} \sum_{r \in R_s^*} a_r x_r^* \qquad (11a)$$

Subject to

$$\sum_{l \in A_f} p_l \sum_{s \in S} x_{sl}^* \geq \sum_{l \in A_f} c_l^* y_l^*, \ \forall f \in Q_F \qquad (11b)$$

$$\sum_{l \in A_{0f}} p_l \sum_{s \in S} x^*_{ls} \geq \sum_{l \in A_f} \sum_{s \in S} m_l x^*_{sl} + \sum_{i \in N_f} q_i v^*_i, \forall f \in Q_M \tag{11c}$$

$$u_s + \sum_{l \in A_r} p_l = U_s + a_r - \left( \sum_{l \in A_r \cap (A_0 \cap A_F \cup A_D \cup A_M)} t_l + \sum_{l \in A_r \cap A_{0M}} \tau_l \left( \sum_{s \in S} x^*_{ls}; h_l \right) \right), \forall r \in R^*_s, \forall s \in S \tag{11d}$$

$$u_s \geq U_s - \left( \sum_{l \in A_{r'} \cap (A_0 \cup A_F \cup A_D \cup A_M)} (t_l + \mu^*_l + c_l(1 - y^*_l)) + \sum_{l \in A_{r'} \cap A_{0M}} \tau_l \left( \sum_{s \in S} x^*_{ls} + 1; h_l \right) \right.$$

$$\left. + \sum_{l \in A_{r'} \cap A_M} m_l + \sum_{i \in N_{r'} \cap N_M} q_i(1 - v^*_i) + \sum_{l \in A_{r'}} p_l \right), \forall r' \in R_s \setminus R^*_s, \forall r \in R^*_s, \forall s \in S \tag{11e}$$

$$u_s \geq 0, \forall s \in S \tag{11f}$$

$$p_l \geq 0, \forall l \in A_r, \forall r \in R^*_s, \forall s \in S \tag{11g}$$

$$a_r \geq 0, \quad \forall r \in R^*_s, \forall s \in S \tag{11h}$$

There are three sets of decision variables: user's payoff of user group $s \in S$ ($u_s$), fare of link $l \in A_r$, $r \in R^*_s$, $s \in S$ ($p_l$), and subsidy to each user on path $r \in R^*_s$, $s \in S$ ($a_r$). Subsidies are injected per path per user to increase trip utilities, which is reflected in the cost allocation constraints (Eq. (11d)). The objective (Eq. (11a)) is to minimize the total amount of subsidies injected. Since subsidy is injected per path per user, path flows are needed for total subsidies in Eq. (11a).

## 4 Proposed solution algorithms

The challenge is in solving for a MaaS platform equilibrium with guaranteed local stability, with or without subsidy. We propose a branch and bound algorithm that can obtain the exact solution to the many-to-many matching model $L_1$ in Section 4.1, but stability is not guaranteed per Lemma 1. For obtaining a solution with guaranteed stability, i.e. the MaaS platform equilibrium, we integrate $L_2$ and $L_3$ as shown in Section 4.2.

### 4.1. Exact solution method to the many-to-many matching model $L_1$

The algorithm is composed of 3 parts nested within each other: branch and bound (Land and Doig, 2010), Lagrangian relaxation with subgradient optimization, and Frank-Wolfe algorithm. An overview of the complete algorithm is presented in Fig. 3 (which also contains parts introduced in 4.2). The integral constraints of $y_{ij}$ and $v_{ih}$ are relaxed in the branch and bound. At each branch, a nonlinear traffic assignment problem with capacity (Eq. (3c)) and MOD node controlling constraints (Eq. (3d)) is solved. The capacity constraints of fixed-route links are removed using a Lagrangian relaxation approach (Algorithm 2). The resulting non-linear traffic assignment problem is solved with a Frank-Wolfe algorithm (Algorithm 3). Feasibility of the original problem in each branch involves checking the binary constraints.

*4.1.1. Branch and bound algorithm*
For each branch, the set of $y_l$ ($v_i$) constrained to be 0 forms set $Y_0$ ($V_0$), the set of $y_l$ ($v_i$) constrained to be 1 forms set $Y_1$ ($V_1$). For each branch, we do the following modification to the network to make sure that constraints (3c) and (3e) are met.

- For $y_l \in Y_0$, the corresponding link $l$ is removed from the network for this branch.
- For $y_l \in Y_1$, all the other links connecting the same nodes owned by the same operator to represent different service frequency options are removed from the network for this branch.
- For $v_i \in V_0$, all the links incident on MOD node $i$ are removed for this branch.
- For $v_i \in V_1$, all the MOD nodes owned by the same operator with a fleet size different from the fleet size associated with node $i$ should be added to $V_0$.

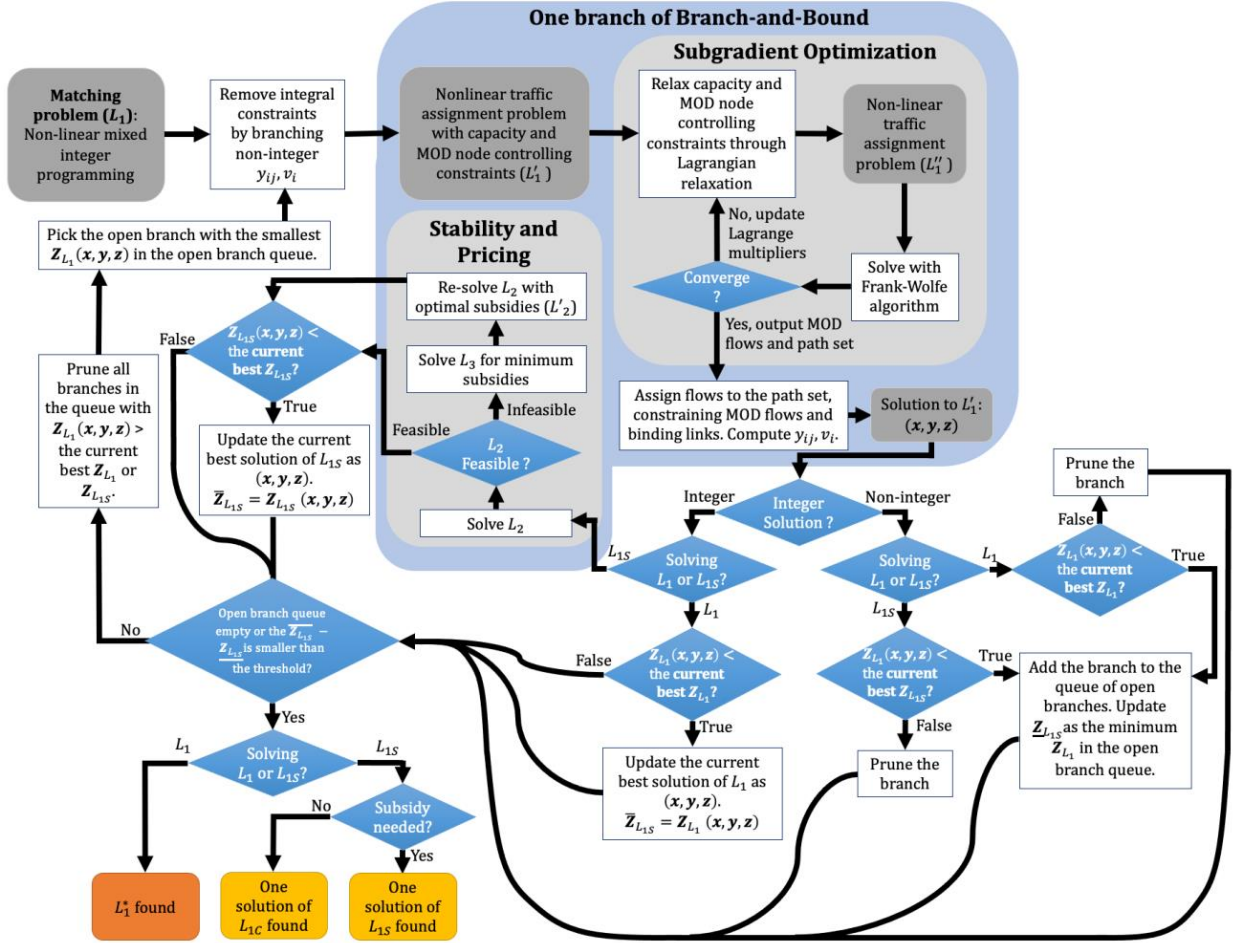

Figure 3. Flow chart of the exact solution algorithm for $L_1$, with added $L_2$ and $L_3$ components for bounded heuristic for $L_{1S}$.

We denote the integral-relaxed $L_1$ of a branch as $B(Y_1, Y_0, V_1, V_0)$. Following the framework of a conventional Branch and Bound algorithm, each $B(Y_1, Y_0, V_1, V_0)$ is solved with Algorithm 1. The solution of $B(Y_1, Y_0, V_1, V_0)$ includes $\hat{X}_R, \hat{X}_L, \hat{Y}, \hat{V}, \hat{Z}_{L_1}, and\ \hat{\mu}$, which denote a vector of path flows on matched paths $r \in R_s^*, s \in S$, a vector of link flows on all links $l \in A$, a vector of values of $y_l$ of fixed-route links $l \in A_F$, a vector of values of $v_i$ of MOD nodes $i \in N_M$, objective function value, and a vector of Lagrange multipliers of the capacity constraints of fixed links $l \in A_F$. We denote the solution set which includes all solution variable of a branch as $\hat{\chi} = [\hat{X}_R, \hat{X}_L, \hat{Y}, \hat{V}, \hat{Z}_{L_1}, \hat{\mu}, Y_1, Y_0, V_1, V_0]$.

For finding a solution to $L_1$, the algorithm stops when all branches are closed. A gap between the upper bound $\overline{Z}_{L_1}$ and lower bound $\underline{Z}_{L_1}$ can be monitored in cases where the maximum run time is exceeded. The property of the branch and bound algorithm ensures that if there exists a feasible integer solution for $L_1$, the algorithm should eventually find it. If the algorithm fails to find any branch with an integer solution, $L_1$ is infeasible.

*4.1.2. Lagrangian Relaxation*

To handle the capacity constraints for the fixed-route links, we adopt the Lagrangian relaxation method from Crainic et al. (2001), which is designed to work with branch and bound algorithms. To improve accuracy, a redundant set of constraints Eq. (12) is added, where $b_{sl} = \min\{d_s, w_{sl}\}$. This set of constraints is named as the strong forcing constraints, with Lagrange multipliers of $\beta_{sl}$.

$$x_{sl} \leq b_{sl} y_l, \quad \forall l \in A_F, \forall s \in S \quad (\beta_{sl}) \quad (12)$$

The strong forcing constraints (12) can be relaxed in a Lagrangian way. The capacity constraints in Eq. (3c) are replaced by Eq. (13) with a Lagrange multiplier of $\gamma_l$. With the replacement, $y_l$ can be computed as Eq. (14), where $c_l(\beta) = c_l - \sum_s b_{sl}\beta_{sl}$.

$$\sum_{s \in S} x_{sl} \leq w_l, \quad \forall l \in A_F \quad (\gamma_l) \quad (13)$$

$$y_l = \begin{cases} \sum_{s \in S} x_{sl} / w_{sl}, & \text{if } c_l(\beta) \geq 0 \\ 1, & \text{otherwise} \end{cases} \quad (14)$$

We relax Eq. (3d) similarly as relaxing Eq. (3c). First, a redundant set of constraints is added as Eq. (15) with Lagrange multipliers $\eta_{sl}$. $A_i^+$ are the outbound links of MOD node $i$, which include MOD links and MOD egress links.

$$x_{sl} \leq d_s v_i, \quad \forall i \in N_M, l \in A_i^+, s \in S \quad (\eta_{sl}) \quad (15)$$

Then Eq. (15) are relaxed in a Lagrangian way. The node controlling constraints in Eq. (3d) are replaced by Eq. (16) with Lagrange multipliers $\lambda_l$. With the replacement, $v_i$ can be computed as Eq. (17), where $q_i(\eta) = q_i - \sum_{s \in S} d_s \sum_{\forall l \in A_i^+} \eta_{sl}$.

$$\sum_{s \in S} x_{sl} \leq \sum_{s \in S} d_s, \quad \forall i \in N_M, l \in A_i^+ \quad (\lambda_l) \quad (16)$$

$$v_i = \begin{cases} \sum_{s \in S} x_{sl} / \sum_{s \in S} d_s, & \text{if } q_i(\eta) \geq 0 \\ 1, & \text{otherwise} \end{cases} \quad (17)$$

The relaxed objective function is shown in Eq. (18). The relaxation of Eq. (3d) leads to cost changes of MOD node outbound links, including MOD links ($A_M$) and MOD egress links. MOD egress links are denoted as $l \in A_E$.

$$\begin{aligned} Z_{L_1}(\gamma, \beta, \lambda, \eta) = &\sum_{s \in S} \sum_{l \in A_O \cup A_D} t_l x_{sl} + \sum_{s \in S} \sum_{l \in A_F} \left( t_l + \gamma_l + {c_l(\beta)^+}/{w_l} + \beta_{sl} \right) x_{sl} \\ &+ \sum_{s \in S} \sum_{l \in A_M} (t_l + m_l + \lambda_l + {q_i(\eta)^+}/{\sum_{s \in S} d_s} + \eta_{sl}) x_{sl} \\ &+ \sum_{s \in S} \sum_{l \in A_E} (\lambda_l + {q_i(\eta)^+}/{\sum_{s \in S} d_s} + \eta_{sl}) x_{sl} + \sum_{s \in S} \sum_{l \in A_{OM}} \tau_l \left( \sum_{s \in S} x_{sl}; h \right) x_{sl} \\ &+ constants \end{aligned} \quad (18)$$

In subgradient optimization, the Lagrange multipliers are updated from 0 when the corresponding constraints are broken. However, constraints Eqs. (15) – (16) will never be broken after removing $v_i$. Hence, $(\lambda_l + q_i(\eta)^+ / \sum_{s \in S} d_s + \eta_{sl})$ will always be $q_i / \sum_{s \in S} d_s$ for all concerned MOD and MOD egress links, which means that the relaxed objective can be rewritten as $L'_1$ in Eq. (19).

$$Z_{L'_1}(\gamma, \beta) = \sum_{s \in S} \sum_{l \in A_O \cup A_D} t_l x_{sl} + \sum_{s \in S} \sum_{l \in A_F} \left( t_l + \gamma_l + c_l(\beta)^+ / w_l + \beta_{sl} \right) x_{sl}$$
$$+ \sum_{s \in S} \sum_{l \in A_M} (t_l + m_l + q_i / \sum_{s \in S} d_s) x_{sl} + \sum_{s \in S} \sum_{l \in A_E} \frac{q_i x_{sl}}{\sum_{s \in S} d_s} \quad (19)$$
$$+ \sum_{s \in S} \sum_{l \in A_{OM}} \tau_l \left( \sum_{s \in S} x_{sl}; h \right) x_{sl} + constants$$

Subgradient optimization of estimating Lagrange multipliers $\gamma_l$ and $\beta_{sl}$ are shown in Algorithm 1. Algorithm 1 is stopped when the change of the Lagrange multipliers is small enough, i.e. when the L2 norm of change of the vector $[\gamma, \beta]$ between 2 consecutive iterations is smaller than a pre-set tolerance $\epsilon$.

**Algorithm 1. Subgradient optimization for each branch**

Initialize path sets $\hat{R}_s = [\ ]$ for all OD pairs $s \in S$. $\hat{R} = [\hat{R}_s, \forall s \in S]$.
Initialize Lagrange multipliers $\gamma_l = 0, \forall l \in A_F$ and $\beta_{sl} = 0, \forall l \in A_F, \forall s \in S$
While the L2 norm of the change of $[\gamma, \beta]$ between 2 consecutive iterations > tolerance $\epsilon$:
    Step size $\theta = 1/$iteration number.
    Solve $L'_1$ using Algorithm 2 with $\gamma_l$ and $\beta_{sl}$, obtaining link flows $F_X$ and updated path set $\hat{R}$.
    Update $\gamma_l$ with $F_X$: $\gamma_l = \max\{0, \gamma_l + \theta(\sum_{s \in S} x^*_{sl} - w_l)\}$.
    Update $\beta_{sl}$ with $F_X$: $\beta_{sl} = \max\{0, \beta_{sl} + \theta(x^*_{sl} - b_{sl})\}$.
Return $\hat{R}, F_X, \gamma, \beta$

### 4.1.3. Modified Frank-Wolfe algorithm

After relaxing capacity and MOD node controlling constraints with Lagrange relaxation, $L_1$ is reduced to a system-optimal traffic assignment problem with congestion effects on MOD access links ($L'_1$), which can be solved with a Frank-Wolfe algorithm (Frank and Wolfe, 1956; LeBlanc et al., 1975) by replacing the average cost functions of MOD access links with marginal cost functions while finding shortest marginal cost paths as shown in Algorithm 2. The algorithm is stopped when $\alpha$ is smaller than a tolerance $\varepsilon$ for $C$ consecutive iterations, where $C$ and $\varepsilon$ are pre-set parameters.

**Algorithm 2. Modified Frank-Wolfe algorithm for solving $L'_1$**

Initialize current link flow set $F_X$: demand of each OD pair all assigned to dummy links.
Count = 0.
For each OD pair $s \in S$:
    Compute link costs of fixed-route links with $\gamma, \beta$: $t'_{sl} = t_l + \gamma_l + c_l(\beta)^+ / w_l + \beta_{sl}$.
    Compute link costs of MOD links: $t'_{sl} = t_l + m_l + q_i / \sum_{s \in S} d_s$.
    Compute link costs of MOD egress links: $t'_{sl} = q_i / \sum_{s \in S} d_s$.
    Compute marginal link costs of MOD access links with current link flows: $t'_{sl} = \frac{\partial \tau_l(\sum_{s \in S} x^*_{sl}; h) \sum_{s \in S} x^*_{sl}}{\partial(\sum_{s \in S} x^*_{sl})}$.
    (Cost of transfer links and walking links are the same as original.)
While Count < $C$:
    Find the shortest path of each OD pair $s \in S$, save the paths found to path sets $\hat{R}_s, s \in S$.
    Assign all demand each OD pair to the shortest paths.

      Aggregate assigned path flows to obtain auxiliary link flows $F_Y$.
      Compute the derivative of Eq. (19) with respect to the vector of $x_{sl}$: $\frac{dO(\gamma,\beta)}{dx}(x)$.
      Solve $\frac{dO(\gamma,\beta)}{dx}(F_X + \alpha(F_Y - F_X))$ for $\alpha$.
      Update $F_X$: $F_X = F_X + \alpha(F_Y - F_X)$.
      If $\alpha <$ tolerance $\varepsilon$:
          Count = Count + 1.
      Else:
          Count = 0
Return $F_X, \hat{R}$.

---

For each branch $B(Y_1, Y_0, V_1, V_0)$, Algorithm 1 obtains the Lagrange multipliers, but not the flows ($\hat{X}_R$, $\hat{X}_L$) directly. Path flows $\hat{X}_R$ need to be obtained with a linear programming assigning demand to the paths in the path set $\hat{R} = [\hat{R}_s, \forall s \in S]$ iteratively updated by Algorithm 2. The path flow solution $\hat{X}_R$ of branch $B(Y_1, Y_0, V_1, V_0)$ can be obtained by solving a linear program minimizing system total adjusted travel cost, where adjusted link costs are the coefficients in Eq. (19). Flows of the MOD links should be the same as the MOD flows output from the final iteration of subgradient optimization, which can be formulated as equality constraints. OD demand is also formulated as equality constraints. Link flows $\hat{X}_L$ can be obtained after having $\hat{X}_R$. Then $\hat{Y}$ and $\hat{V}$ can be obtained using Eq. (14) and Eq. (17). With $\hat{X}_L, \hat{Y}$ and $\hat{V}$ known, the corresponding equilibrium objective value $\hat{Z}_{L_1}$ can be computed using Eq. (3a).

**Proposition 4**. Exact solution algorithm for $L_1$. *The proposed branch and bound algorithm with branching shown in Fig. 3 is an exact algorithm for finding an optimal solution to $L_1$.*

***Proof***. The branch and bound relaxes the integral constraints of variables $y_l$ and $v_i$. Only branches whose solution satisfies the integral constraints are considered feasible. These feasible solutions are used in updates of the upper bound $\overline{Z}_{L_1}$. Since the nonlinear program after relaxing the set of binary variables is convex and the combinations of binary variables are finite, the Lagrange relaxation and Frank-Wolfe portions will converge to exact upper bounds. ∎

4.2. Bounded heuristic for a locally stable solution
Due to Lemma 1, finding a MaaS platform equilibrium requires finding a stable solution (i.e. solving $L_{1C}$). We propose integrating $L_2$ and $L_3$ into the exact solution algorithm above that guarantees a locally stable MaaS platform equilibrium that may require subsidy as shown in Fig. 3 with the "Stability and Pricing" component.

**Proposition 5**. *The proposed heuristic in Fig. 3 guarantees a locally stable solution that may or may not require subsidy.*

***Proof***. A stable integral solution found using $L_2$ without subsidy means that a locally stable solution ($\tilde{Z}_{L_{1C}}$) is found, where $\tilde{Z}_{L_{1C}} \geq Z^*_{L_{1C}}$. The globally stable MaaS platform equilibrium ($Z^*_{L_{1C}}$) is not ensured to be found. The solution to $L_{1C}$ may be prematurely pruned in the branch and bound process if the solution is under an integral branch for $L_1$. However, if the optimal solution of $L_1$ is also locally stable, it is ensured as the MaaS platform equilibrium since no stable solution with lower objective value exists.
    In the same manner, the optimal $Z^*_{L_{1S}}$ may be prematurely pruned, but when all branches are closed we are guaranteed to have a feasible solution to a subsidized MaaS platform equilibrium. We denote this solution as $\tilde{Z}_{L_{1S}}$, where $\tilde{Z}_{L_{1S}} \geq Z^*_{L_{1S}}$. In other words, the algorithm in Fig. 3 is guaranteed to obtain either $\tilde{Z}_{L_{1S}}$ (requiring subsidy) or $\tilde{Z}_{L_{1C}}$ (not requiring subsidy), whichever performs better. ∎

Two sets of constraints are relaxed in the branch and bound: the integral constraints of $y_l$ and $v_i$, and stability constraints. Algorithm 3 shows the process of solving one branch of branch and bound integrating $L_2$ and $L_3$. We denote the integral-stability-relaxed problem of a branch as $B(Y_1,Y_0,V_1,V_0)$, which can be solved with Algorithm 1. When an integral solution is found after solving $B(Y_1,Y_0,V_1,V_0)$, we solve the cost allocation problem ($L_2$) for the corresponding matching. If $L_2$ is feasible, the stable outcome space is non-empty. With a non-empty stable outcome space and an $L_1$ objective value lower than the lowest subsidized system cost found yet, the upper bound the $\overline{Z}_{L_{1S}}$ is updated and a locally stable solution without subsidy ($\tilde{Z}_{L_{1C}}$) is found.

If an empty stable outcome space is reached, we solve the minimum subsidy problem ($L_3$) to find the subsidized system cost ($Z_{L_{1S}}$). If the subsidized system cost is lower than the lowest subsidized system cost found yet, the upper bound the $\overline{Z}_{L_{1S}}$ needs updating. The complete heuristic for $L_{1S}$ is shown in Fig. 3. When all branches are pruned, the current upper bound the $\overline{Z}_{L_{1S}}$ (and corresponding stable outcome space) is output as $\tilde{Z}_{L_{1S}}$.

**Algorithm 3. One branch of branch and bound for bounded heuristic**

Inputs: sets of integer variables constrained to be 1 and 0: $Y_1, Y_0, V_1, V_0$; solution of the branch with an integral solution and the lowest subsidized system cost found yet: $\tilde{X}_L, \tilde{X}_R, \tilde{Y}, \tilde{V}, \tilde{\mu}, \tilde{U}, \tilde{P}, \tilde{\mathcal{A}}, \tilde{Z}_{L_{1S}}, \mathbb{Q}$.
Solve $B(Y_1,Y_0,V_1,V_0)$ with Algorithm 1 to obtain $\hat{\chi} = [\hat{X}_R, \hat{X}_L, \hat{Y}, \hat{V}, \hat{Z}_{L_1}, \hat{\mu}, Y_1, Y_0, V_1, V_0]$.
If $\hat{Y}, \hat{V}$ are all integers:
    Solve $L_2(\hat{\chi})$ to obtain $\hat{U}, \hat{P}$.
    If $L_2(\hat{\chi})$ is feasible:
        $\hat{\mathcal{A}} = [\,]$. $\hat{Z}_{L_{1S}} = \hat{Z}_{L_1}$.          # A locally stable solution found
        Add $\hat{\chi}$ to the set of locally stable solutions.
    Else:
        Solve $L_3(\hat{\chi})$ to obtain $\hat{\mathcal{A}}$. Total subsidy is $\hat{X}_R\hat{\mathcal{A}}$.
        Solve $L'_2(\hat{\chi}, \widehat{\mathcal{A}})$ to obtain $\hat{U}, \hat{P}$.
        $\hat{Z}_{L_{1S}} = \hat{Z}_{L_1} + \hat{X}_R\hat{\mathcal{A}}$.
    If $\hat{Z}_{L_{1S}} < \tilde{Z}_{L_{1S}}$:
        $\tilde{X}_R = \hat{X}_R, \tilde{X}_L = \hat{X}_L, \tilde{Y} = \hat{Y}, \tilde{V} = \hat{V}, \tilde{Z}_{L_{1S}} = \hat{Z}_{L_1}, \tilde{\mu} = \hat{\mu}, \tilde{U} = \hat{U}, \tilde{P} = \hat{P}, \tilde{\mathcal{A}} = \hat{\mathcal{A}}$.
Else:
    If $\hat{Z}_{L_1} < \tilde{Z}_{L_{1S}}$:
        Append $\hat{\chi}$ to $\mathbb{Q}$.
Return $\tilde{X}_L, \tilde{X}_R, \tilde{Y}, \tilde{V}, \tilde{\mu}, \tilde{U}, \tilde{P}, \tilde{\mathcal{A}}, \tilde{Z}_{L_{1S}}, \mathbb{Q}$, the set of locally stable solutions.
# Updated current best Subsidized MaaS platform equilibrium

We denote a cost allocation problem given an integral solution $\hat{\chi}$ of $B(Y_1,Y_0,V_1,V_0)$ as $L_2(\hat{\chi})$. The solution of $L_2(\hat{\chi})$ includes $\hat{U}$ and $\hat{P}$ (buyer optimal or seller optimal or both), which denote a vector of users' payoffs of users $s \in S$ and a vector of link fares of matched links $l \in A_r$, $r \in R^*_s, s \in S$. We denote a subsidy problem given an integral solution $\hat{\chi}$ of $B(Y_1,Y_0,V_1,V_0)$ as $L_3(\hat{\chi})$. The solution of $L_3$ is $\hat{\mathcal{A}}$, which denotes a vector of optimal subsidies on matched paths $r \in R^*_s, s \in S$. We denote the cost allocation with the solved subsidies as $L'_2(\hat{\chi}, \hat{\mathcal{A}})$. Solution of $L'_2$ also includes $\hat{U}$ and $\hat{P}$.

We show that the solution has a worst-case bound defined by the subsidized equilibrium objective of the matching solution for $L_1$ (Corollary 2).

**Corollary 2**. Worst-case bound. *The heuristic for $L_{1S}$ in Fig. 3 has an objective value $\tilde{Z}_{L_{1S}}$ bounded from above, i.e. $\tilde{Z}_{L_{1S}} \leq Z^*_{L_1}(x,y,v) + Z^*_{L_3|(x,y,v)}$.*

*Proof.* The bound used in the heuristic for $L_{1S}$ ($Z_{L_1}(x,y,v) + Z_{L_3|(x,y,v)}$) is greater than or equal to the bound used in the exact algorithm for $L_1$ ($Z_{L_1}(x,y,v)$). The larger bound still ensures that the optimal solution of $L_1$ is found in the branch and bound algorithm. Hence the final solution $\tilde{Z}_{L_{1S}}$ found by the heuristic is ensured to have a subsidized system cost smaller than the optimal solution of $L_1$, i.e. $\tilde{Z}_{L_{1S}} \leq Z^*_{L_1}(x,y,v) + Z^*_{L_3|(x,y,v)}$. ∎

The difference between the exact solution to $L_1$ and the heuristic for $L_{1S}$ is the upper bound used in the branch and bound, which are $\overline{Z}_{L_1}$ and $\overline{Z}_{L_{1S}}$. The difference between $\overline{Z}_{L_1}$ and $\overline{Z}_{L_{1S}}$ is the subsidy. Note that there is not a meaningful bound of the amount of subsidy needed to stabilize the game that we can prove, due to the complicated dependency between the subsidy and the network structure, link ownership, and assignment. If all the integral solutions in the branch and bound algorithm are stable without subsidy (non-empty), $\overline{Z}_{L_1} = \overline{Z}_{L_{1S}}$, indicating that the final solution found is also the optimal solution to $L_1$ and the MaaS platform equilibrium ($Z^*_{L_{1C}}$) (Corollary 3).

**Corollary 3**. Optimal condition. *The heuristic for $L_{1S}$ in Fig. 3 leads to the optimal solution to $L_1$ and the MaaS platform equilibrium ($Z^*_{L_1} = Z^*_{L_{1C}}$) if all the integral solutions found by the branch and bound are stable without subsidy (non-empty).*

***Proof***. The upper bound used in the heuristic for $L_{1S}$ ($\overline{Z}_{L_{1S}} = Z_{L_1}(x,y,v) + Z_{L_3|(x,y,v)}$) is greater than or equal to the upper bound used in the exact algorithm for $L_1$ ($\overline{Z}_{L_1} = Z_{L_1}(x,y,v)$) due to the subsidy injected to unstable integer solutions. When all the integer solutions found by the branch and bound are stable without subsidy (non-empty), $\overline{Z}_{L_1} = \overline{Z}_{L_{1S}}$. In this case, the solution found by the heuristic is also the optimal solution to $L_1$. Since the optimal solution to $L_1$ is also stable without subsidy, there is no other stable solution with a lower objective value, indicating that the optimal solution to $L_1$ is also the MaaS platform equilibrium ($Z^*_{L_1} = Z^*_{L_{1C}}$). ∎

## 5 Numerical experiments

Two sets of computational experiments are conducted to illustrate the heuristic of finding a guaranteed locally stable solution (a solution to $L_{1S}$ or $L_{1C}$). The first is used to verify the methodology and illustrate the differences between locally and globally stable solutions. The second, tested on an expanded version of Sioux Falls, illustrates generalizable insights that can be gained from the use of this model.

### 5.1 Small Illustrative Case

We use a toy network shown in Fig. 4 to illustrate how the method works. The original network is shown in Fig. 4(a). All costs are in dollars ($). The solid links represent the fixed-route services, links with the same color are operated by the same operator. Link (21,22) and (21,23) are transfer links between lines, which are without capacity and operating cost with no owners. There are 3 MOD operators (blue, green, brown). The circles represent the service zones that MOD operators can choose from to operate (blue: A,B,C; green: B,C; brown: B,D). Zone A covers transit station node 1. Zone B covers transit station nodes 21,22,23. Zone C covers transit station node 3. Zone D covers transit station node 4.

The network is expanded into Fig. 4(b) by creating complete subgraphs for each MOD operator and adding MOD access links and egress links. Travel cost, operating cost, and capacities are labelled as shown in the legend. The access/wait cost functions of all the MOD operators on all MOD access links $l$ are $\tau_l(\sum_{s \in S} x_{sl}; h) = 0.5 h^{-2} \sum_{s \in S} x_{sl}$. The MOD operating cost parameter of all the MOD operators at all MOD nodes is $m_l = h^2$. Fleet size choices of all MOD operators are 1, 2, and 3. The network is further expanded to Fig. 4(c) to represent the 3 fleet size options. Infrastructure cost of MOD nodes 7, 8, 9, 10, 11, 12, and 13 are 3, 3, 2, 2, 1, 1, and 3, respectively for all fleet size options. Demand is 1,000 from node 1 to 3, and 500 from node 1 to 4. Trip utility $U_s$ is $9.50 for both OD pairs. The tolerance $\epsilon$ for subgradient

optimization is 0.05. The tolerance $\varepsilon$ of Frank-Wolfe is 0.01 and the required consecutive number of iterations meeting the tolerance is 5. No optimality gap control is applied, the algorithm is terminated when all branches are pruned.

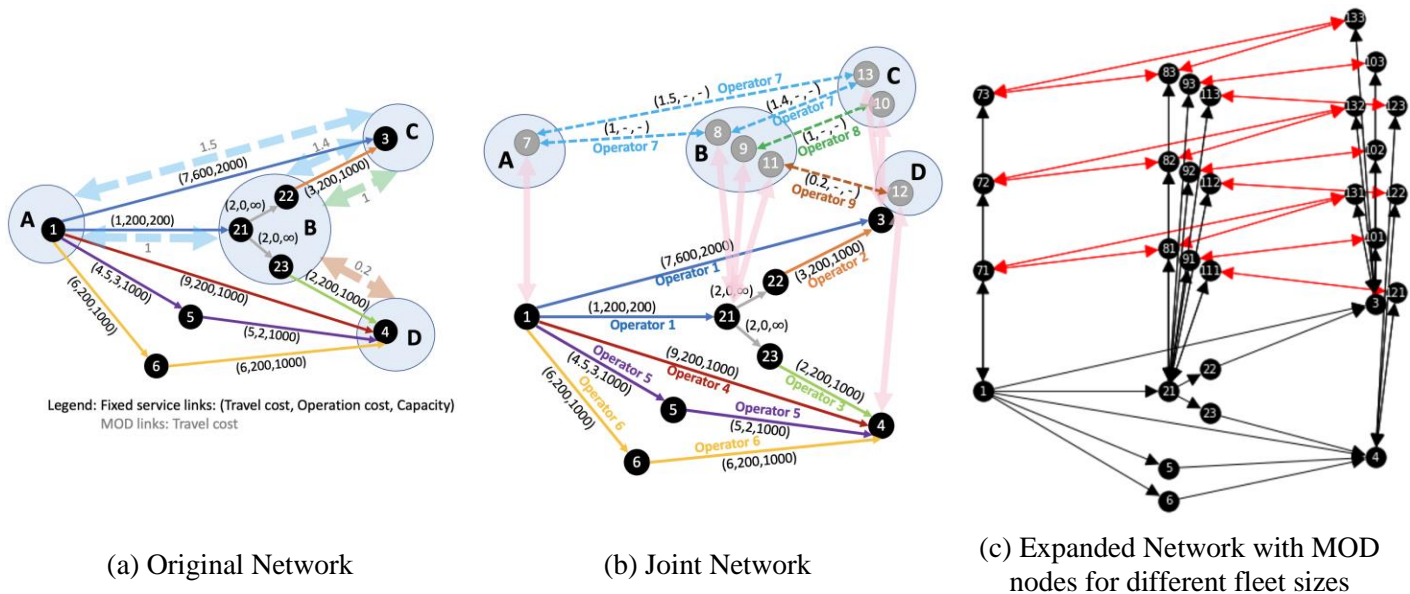

(a) Original Network   (b) Joint Network   (c) Expanded Network with MOD nodes for different fleet sizes

Figure 4. Toy network.

The branch and bound algorithm is coded in Python 3.8.5. It converged after running 25 branches in 11min 47sec (28 sec per branch on average) using a laptop with 2.3 GHz Quad-Core Intel Core i7. One locally stable solution is found without subsidy. The system total cost is $Z_{L_1}^* = Z_{L_1C} = 11,842.87$. Flows and corresponding fares of the solution are shown in Fig. 5. Both MOD Operator 7 and 9 chose to operate at fleet size 1.

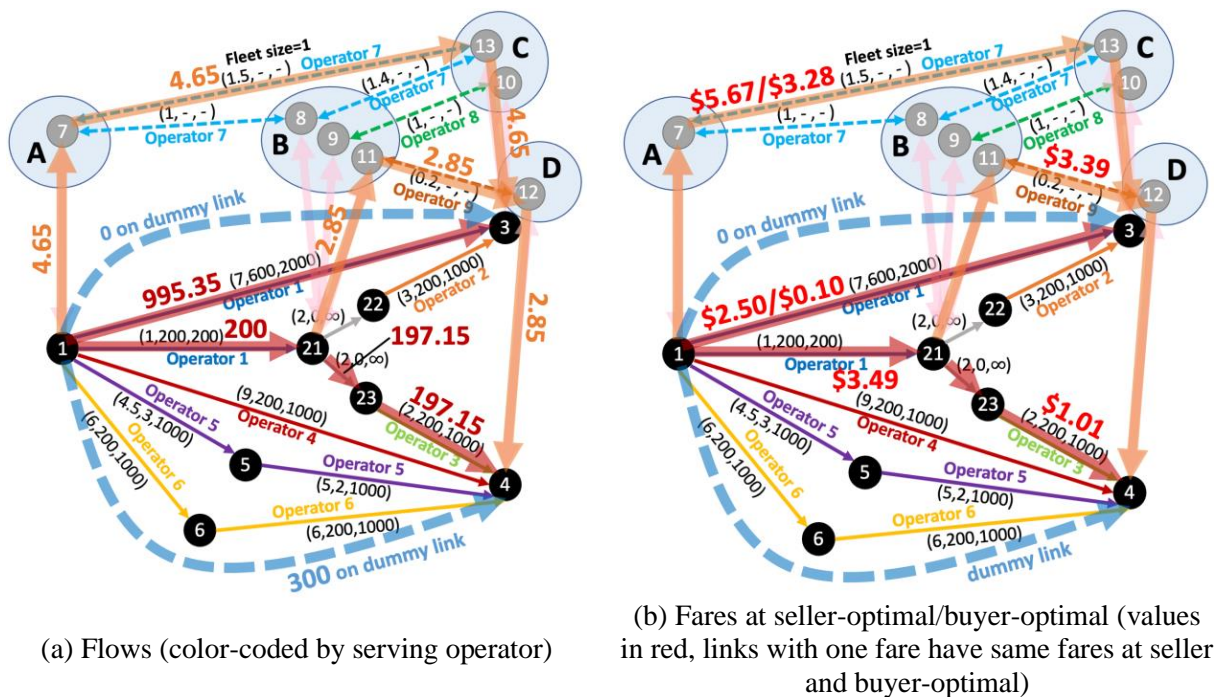

(a) Flows (color-coded by serving operator)   (b) Fares at seller-optimal/buyer-optimal (values in red, links with one fare have same fares at seller and buyer-optimal)

Figure 5. (a) Optimal flows, fleet sizes, and (b) fares of the MaaS platform equilibrium.

In the MaaS platform equilibrium shown in Fig. 5, fixed-route Operators 2, 4, and 6, and MOD Operator 8 do not enter this market. There are travelers choosing not to use the platform (300 on dummy link from 1 to 4). Demand using MOD is small. MOD Operator 7 has a flow of 4.65 going from 1 to 13. MOD operator 9 has a flow of 2.85 going from 21 to 4. Operators 1 and 3 serve the majority of travelers. Links with the same fares for buyer and seller optimal ((1,21),(23,4),(11,12)) are on paths with no extra payoff to allocate in addition to paying for operating cost (path [1,21,23,4], [1,21,11,12,4]). Operator revenues under seller-optimal and buyer-optimal allocations are shown in Table 2.

Table 2. Operator revenue under seller-optimal and buyer-optimal allocations

| Operator ID | Optimal solution, stable without subsidy ($Z_{L_{1C},3} = 11{,}842.87$) | |
| --- | --- | --- |
| | Buyer-optimal | Seller-optimal |
| 1 | $800 | $ 3185.47 |
| 2 | Not Operating | |
| 3 | $200 | $200 |
| 4 | Not Operating | |
| 5 | Not Operating | |
| 6 | Not Operating | |
| 7 | $15.25 | $ 26.40 |
| 8 | Not Operating | |
| 9 | $9.66 | $ 9.66 |

## 5.2 Sioux Falls Case

The algorithm is further tested on the Sioux Falls network. The traditional Sioux Falls network is modeled as a combination of a walking/transfer network and fixed-route transit service segments. Network parameters (link costs and capacities) are shown in Appendix A. Walking links have 0 operating cost and are not owned by any operator. All fixed transit links have an operating cost of $400. No alternative capacities are modeled for the fixed-route services. There are 4 fixed transit lines in this case, marked in Fig. 6 with blue (Operator 1), pink (Operator 2), yellow (Operator 3), and green (Operator 4). All other links are walking/transfer links. The four fixed-route operators can choose to operate or not on each of the links they own. Three MOD operators operate in the region, marked purple (Operator 5), light blue (Operator 6), and orange (Operator 7). Candidate service regions that they cover are marked in Fig. 6. The cost of opening each MOD node is $10, $5, and $15 for MOD operator 5, 6, and 7, respectively. Alternative fleet sizes for all 3 MOD operators are 1 and 2. OD demand is shown in Appendix B. The OD pairs are all the combinations between nodes 1, 2, 12, 18, 13, and 20 (30 OD pairs), totaling 9,700 trip demand. Utility $U_s$ is $20 for all OD pairs. Note that while different values can be specified for different OD pairs, we assume the same trip purpose for all the OD pairs in this experiment, such that volumes of trips at longer distances would be smaller than trips at smaller distances (aligned with conventional gravity model-based trip distribution assumptions). The access/wait cost function and operating cost function of MOD are the same as the illustrative case. The access/wait cost function is $\tau_l(\sum_{s \in S} x_{sl}; h) = 2h^{-2} \sum_{s \in S} x_{sl}$ while the MOD link operating cost function is $m_l = 4h^2$.

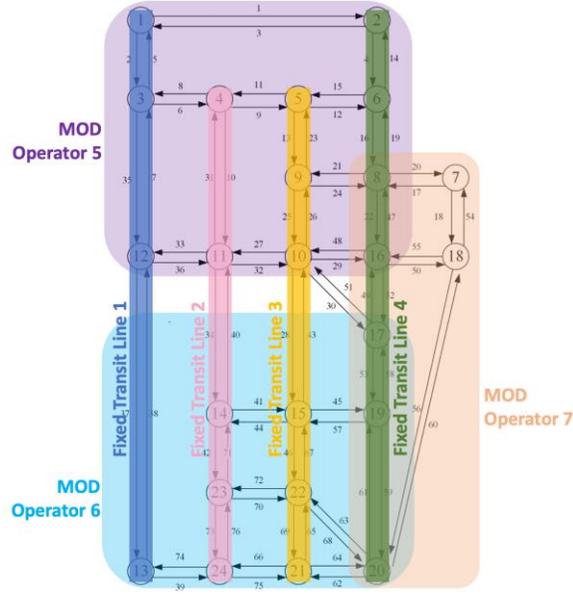

Figure 6. Network construction of the Sioux Falls case.

The expanded network has 82 nodes and 748 links. All the nodes covered by each MOD operator are mutually connected with MOD links as complete subgraphs. The travel costs of MOD links are set to be 75% of the cost of the shortest path between the 2 nodes on the original network. MOD nodes and complete subgraphs are replicated to represent fleet sizes 1 and 2. Corresponding MOD nodes and original nodes are connected with MOD access links and egress links. The tolerance $\epsilon$ for subgradient optimization is 0.05. The tolerance $\varepsilon$ of Frank-Wolfe and the required consecutive number of iterations is 0.01 and 5. All cases are run on a laptop with 2.3 GHz Quad-Core Intel Core i7 and 32 GB 3733 MHz LPDDR4X memory.

*5.2.1. Base Case results*
The proposed algorithm converges after solving 157 branches, taking 4h 1min (1min 32sec per branch on average). Flows and fares of the final solution are shown in Fig. 7. The final solution is the only solution found in the by the branch and bound, hence the optimal solution to $L_1$ (i.e. $Z^*_{L_1}$). The final solution is stable without subsidy, indicating that it is a MaaS platform equilibrium (i.e. $Z_{L_1c}$) according to Corollary 3. The objective value is $108,800.

Even if the MOD links are 25% faster than the shortest path of non-MOD links, there is no MOD operation due to the high operating and infrastructure costs. Within the 4 transit lines, only the blue line operated by Operator 1 ended up operating. Users who travel on the horizontal direction walk to the stations of the blue line (node 1,3,12,13) and take the blue line. Operator revenue, users' payoff, and unserved demand are shown in the first column of Table 3. Unserved demand is 1,200 (12.4% of total demand). These users are not able to obtain non-negative payoffs given the current market and built environment.

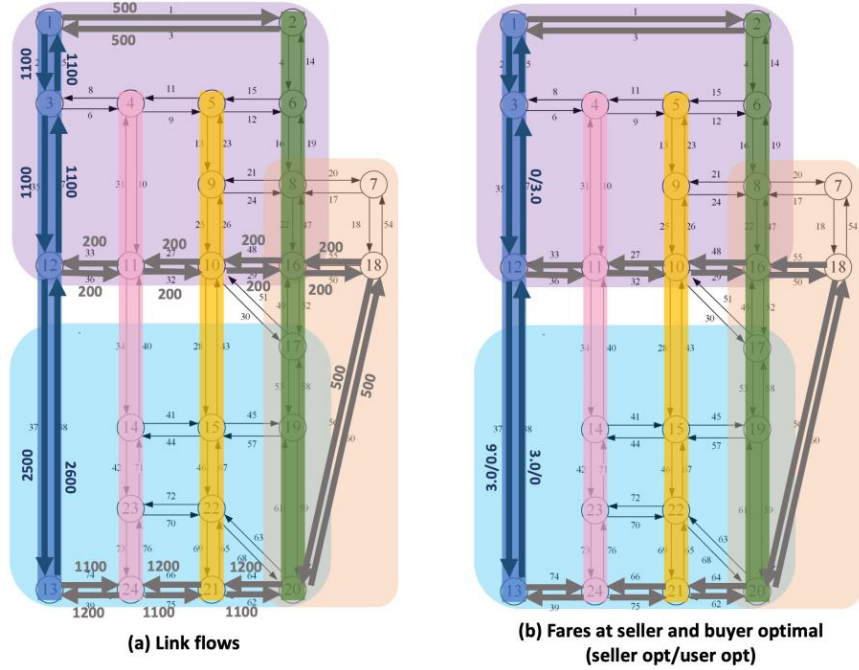

Figure 7. Base case assignment and fares (links with one fare have same fares at seller and buyer-optimal, links with no fares have 0 fares at both seller and buyer-optimal).

### 5.2.2. Reducing MOD Operating cost

We model a scenario where the MOD access and operating costs are further reduced for Operator 5 (purple) by 50% to $\tau_l = 2h^{-2}\sum_{s \in S} x_{sl}$ and $m_l = 2h^2$. Lower MOD operating costs could mean advances in MOD service technologies. For example, the reduction may be from adoption of better routing and matching algorithms and more efficient rebalancing strategies.

In this scenario, the proposed algorithm converges after 339 branches, taking 10h 23min (86 sec per branch on average). The final solution requires subsidy. The subsidized system cost is $108,781.08. One other subsidized MaaS platform equilibrium is found, whose subsidized system cost ($108,811.32) is higher than the final solution. No locally stable solution without subsidy is found. The flows and fares of the final solution are shown in Fig. 8, in which solid lines represent non-MOD flows and dashed links represent MOD flows.

Table 3. Seller- and Buyer-optimal operator revenue for the scenarios (percentage changes relative to base case)

| Cases | | Base Case | | Reducing MOD Access and Operating cost (Operator 5 Reduced 50%) | | Heterogeneous Demand | | Heterogeneous Demand with Optimality Gap Control | |
|---|---|---|---|---|---|---|---|---|---|
| Vertex of Stable Outcome Space | | Seller-optimal | Buyer-optimal | Seller-optimal | Buyer-optimal | Seller-optimal | Buyer-optimal | Seller-optimal | Buyer-optimal |
| Operator Revenue | 1 (Blue line) | $15300 | $4800 | $11098.64 | $4800 | $21750 | $4800 | $21750 | $4800 |
| | 2 (Pink line) | Not Operating | | Not Operating | | Not Operating | | Not Operating | |
| | 3 (Yellow line) | Not Operating | | Not Operating | | Not Operating | | Not Operating | |
| | 4 (Green line) | Not Operating | | Not Operating | | Not Operating | | Not Operating | |
| | 5 (Purple MOD) | Not Operating | | $94.71 | $66.25 | $56.08 | $39.52 | $60.22 | $50.44 |
| | 6 (Blue MOD) | Not Operating | | $5.17 | $5.17 | $5.11 | $5.11 | $12.53 | $12.53 |

| 7 (Orange MOD) | Not Operating | | Not Operating | | Not Operating | | Not Operating | |
|---|---|---|---|---|---|---|---|---|
| Total | $15300 | $4800 | $11198.53 (-26.8%) | $4871.43 (+1.4%) | $21811.19 (+42.6%) | $4844.63 (+0.9%) | $21822.76 (+42.6%) | $4862.97 (+1.3%) |
| Fixed transit total | $15300 | $4800 | $11098.64 (-27.5%) | $4800 | $21750 (+42.2%) | $4800 | $21750 (+42.2%) | $4800 |
| MOD total | - | - | $99.88 | $71.43 | $61.19 | $44.63 | $72.76 | $62.97 |
| Users' total payoff (within platform) | $74700 | $85200 | $78900 (+5.6%) | $85227.10 (+0.03%) | $69051.04 (-7.6%) | $86017.60 (+1.0%) | $69050 (-7.6%) | $86017.99 (+1.0%) |
| Total subsidy needed | $0 | | $4.09 | | $3.94 | | $4.45 | |
| Subsidized system cost | $108,800 | | $108,781.08 | | $107990.28 | | $107999.13 | |
| Total unserved demand | 1200 | | 1188.42 (↓0.97%) | | 2243.87 (+87.0%) (High income: 593.87, Low income: 1650) | | 2244.04 (+87.0%) (High income: 594.04, Low income: 1650) | |
| Run time | 4h 1min | | 10h 23min | | 12h 33min | | 10h 13min | |
| Optimality Gap | 0 | | 0 | | 0 | | $13.14 | |

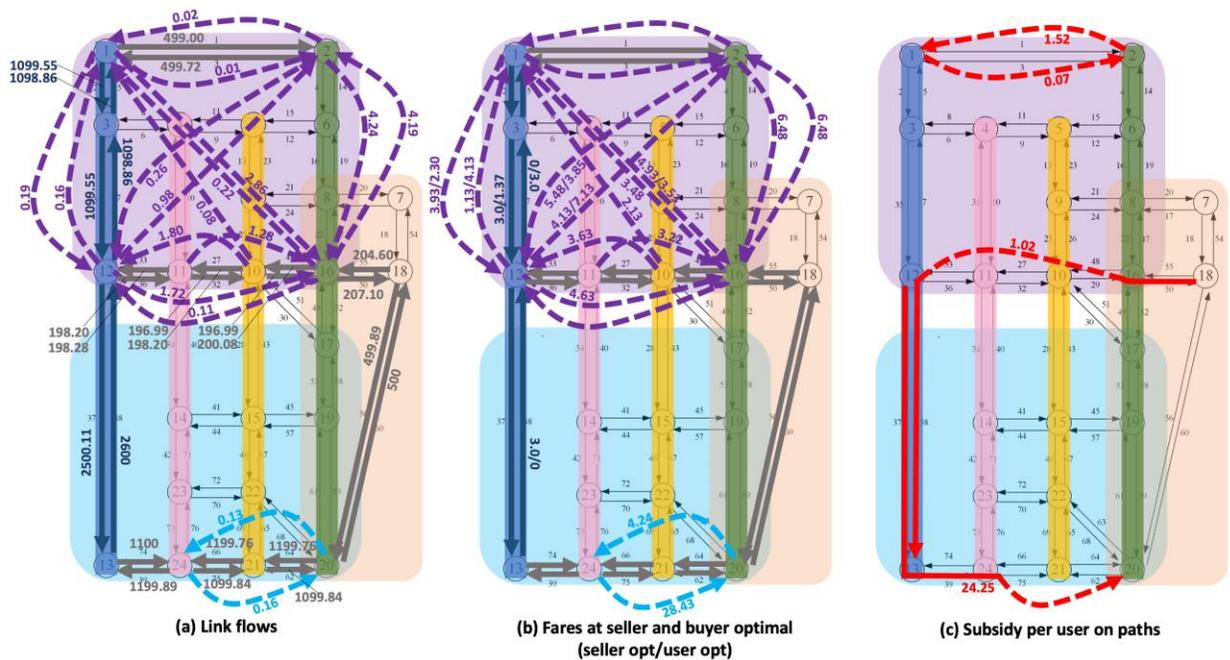

Figure 8. Operator 5 access and operating cost reduced by 50% (links with one fare have same fares at seller and buyer-optimal, links with no fares have 0 fares at both seller and buyer-optimal).

The lower cost of Operator 5 leads to market entry for Operator 5 to serve zones (1, 2, 3, 8, 10, 11, 12, 16) with fleet size 1. Because subsidy is needed, the outcome space is a single solution representing both buyer- and seller-optimal outcomes. Subsidy is required to stabilized path [(1,2)(MOD)], [(2,1)(MOD)], [(18,16),(16,12)(MOD),(12,13)(transit)], and [(12,13)(transit),(13,24),(24,20)(MOD)] which are all paths with MOD. The sum of fare and travel costs of the 4 paths are $6.07, $7.52, $18.02, and $40.25. In the 4 paths, only [(12,13)(transit),(13,24),(24,20)(MOD)] exceeds trip utility. The 4 paths are subsidized since there exist shorter paths (considering the sum of travel costs and fares) that connect the same OD pairs that are assigned flows. The shorter paths for the 4 OD pairs are [(1,2)], [(2,1)], [(18,20),(20,21),(21,24),(24,13)], and [(12,13)(transit),(13,24),(24,21),(21,20)], whose total costs (sum of travel costs and fares) are $6, $6, $17, and $16. The subsidies balance out the differences in path costs.

Although the access and operating cost reduction is only for Operator 5, the subsidy induces market entry for Operator 6 (blue MOD) with a fleet size of 1 (operating between 20 and 24). This is because transit line 1 (blue line) changes on link (3,12) instead of (12,13), making the path [(12,13)(transit),(13,24),(24,20)(MOD)] cheaper and feasible with $ 24.25 of subsidy. The subsidy is large, but the subsidized system total cost is still lower than the base case due to the cost reduction of Operator 5. With the subsidy injected, path [(20,24)(MOD),(24,13),(13,12)(transit)] also appears since it makes more sense for Operator 6 to provide bidirectional service with the nodes 20,24 opened. Hence, the benefit of cost reduction of Operator 5 is shared by Operator 5 and 6, which reflects coopetition between operators in an indirect way (Remark 1). In other cases, such coopetition can be reflected in a more direct manner. For example, with a multimodal path operated by both Operator 5 and 6, cost reduction of Operator 5 could directly lead to more room of profit for Operator 6 on this path, due to their cooperation of operating the path.

**Remark 1**. *The proposed model can quantify changes in operating costs for one operator in a MaaS platform inducing other operators' market entry decisions due to coopetitive effects.*

Operator 5 serves nodes 10 and 11 even though no demand directly goes there. The reason is the high congestion effects at nodes 12 and 16 that push some travelers to walk to 10 and 11 to access MOD service (Remark 2).

**Remark 2**. *The proposed model can quantify impacts of congestion on operators' design decisions.*

Operator revenue, users' payoff, and unserved demand are shown in the second column of Table 3. Total unserved demand dropped slightly to 1188.42 (12.3% of total demand). In this case, the impact of a lower operating cost for Operator 5 on total supply of the system is not significant. Further cost reduction might lead to less unserved demand.

More MOD operation also leads to higher computation time for each branch. With the same tolerance, Lagrange multipliers converge more slowly with MOD access links involved, since path costs are adjusted each iteration due to MOD flow changes.

*5.2.3. Heterogeneous Demand*

In real-life cases, users are heterogeneous. We model two different income groups by assigning different trip utilities. Demand of every OD pair are evenly split into 2 halves. The one with higher income has a trip utility of $U_s = \$24$ while the one with lower income has a trip utility of $U_s = \$16$, resulting in the same population average trip utility as the base case of $20.

The proposed algorithm converged after 205 branches, taking 12h 33min (3min 40sec per branch on average). Computation time is larger due to larger dimensionality of user groups, which is increased from 30 to 60 to represent the two income levels. The final solution is not locally stable without subsidy, which means that it is a solution to $L_{1S}$, i.e. a feasible solution to the subsidized MaaS platform equilibrium. The subsidized system cost is $107,990.13. There are 1 other subsidized solution and 1 stable solution without subsidy found in branch and bound, with system costs higher than the final solution.

The solution is shown in Fig. 9, in which solid lines represent non-MOD flows and dashed links represent MOD flows. Cost allocation results are shown in the third column of Table 3. Compared with the base case, MOD Operators 5 and 6 enter the market (nodes 2, 3, 16, 20, 24) both with fleet size of 1, even if the cost parameters of MOD are the same as the base case. The difference is that only higher-income users are served by MOD services. Unserved demand substantially increases from the base case to 2243.87, in which 1,650 (73.5%) are lower-income users. In other words, by simply having a heterogeneous population, we reveal inequities as more mobility services enter the market but only serve users with higher income levels (Remark 4).

**Remark 4**. *The modeling of heterogeneous user groups can capture inequities in the market where operators enter primarily to serve higher income users while leaving more lower income users out of the platform.*

Subsidy is required to stabilize path [(12,13)(transit),(13,24),(24,20)(MOD)] ($27.05 per unit demand), which serves the higher-income group. The sum of fare and travel cost of a user on the path is $48.05. The cost exceeds the cost of path [(12,13),(13,24),(24,21),(21,20)] ($21) which connects the same OD pair by $27.05, which is the amount of subsidy needed. MOD fares at buyer-optimal and seller-optimal vertices are the same for some links given very narrow stable outcome space of the paths involved. For links on subsidized paths, the subsidies are injected to ensure minimum payoff that satisfies profitability of operators involved, which also leads to same fares at buyer-optimal and seller-optimal vertices.

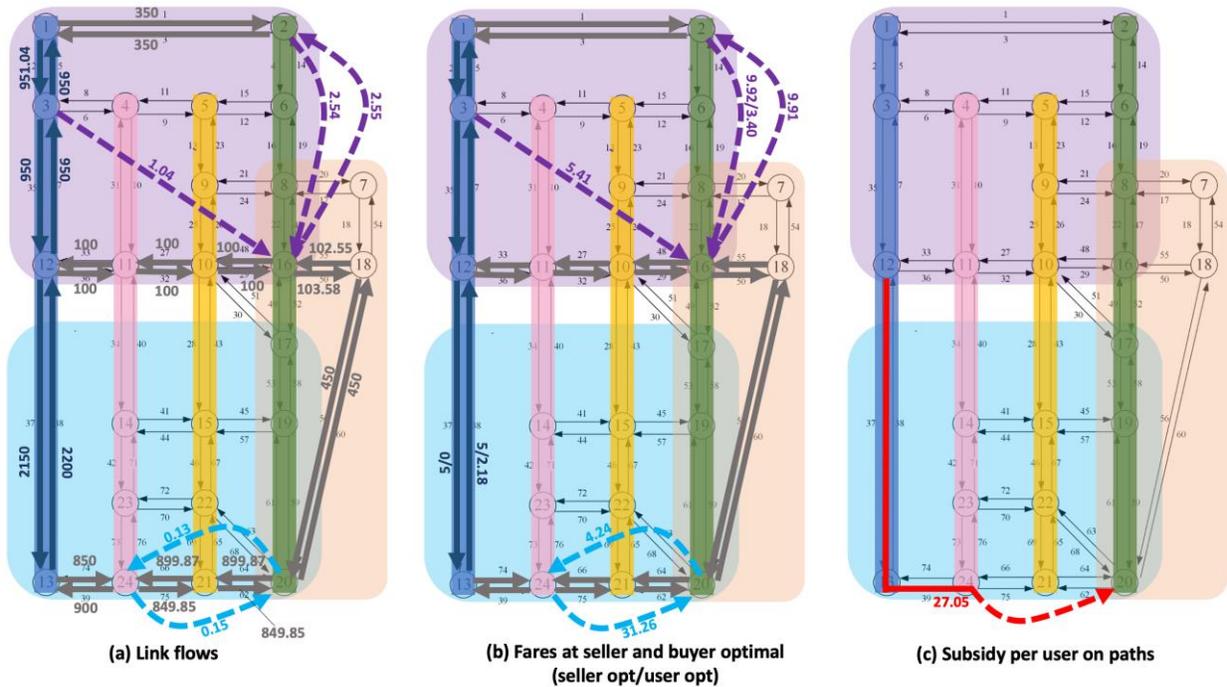

Figure 9. Heterogeneous Demand Case (links with one fare have same fares at seller and buyer-optimal, links with no fares have 0 fares at both seller and buyer-optimal).

*5.2.4. Heterogeneous Demand Case with Optimality Gap Control*

In this case, we test how the optimality gap threshold in the Branch and Bound algorithm ($\overline{Z}_{L_1} - \underline{Z}_{L_1}$) controls the run time and how it affects the accuracy of the solution. We set the optimality gap threshold to be 5% and test it with the heterogeneous demand case with the same settings as Section 5.2.4. The proposed algorithm converged after 197 branches, taking 10h 13min (3min 6sec per branch on average). The final solution is not locally stable without subsidy, which means that it is a solution to $L_{1S}$, i.e. a feasible solution to the subsidized MaaS platform equilibrium. The subsidized system cost is $107,999.13. This is the first subsidized stable solution found in branch and bound. The upper bound of subsidized system cost with the satisfied threshold is $107,999.12 and the lower bound is $107,985.98. The convergence gap is $13.14, which is 0.01% of the lower bound. The run time is reduced by 18.6% of the run time without the optimality gap control. This case highlights the impact of optimality gap control on reducing run time.

The solution is shown in Fig. 10, in which solid lines represent non-MOD flows and dashed links represent MOD flows. The flows and operations are slightly different from the solution without optimality gap control. Flows on the links with major flows have very small differences (blue line; walking links connecting 12,11,10,18; walking links connecting 1,2; walking links connecting 13,24,21,20; walking links connecting 18,20).

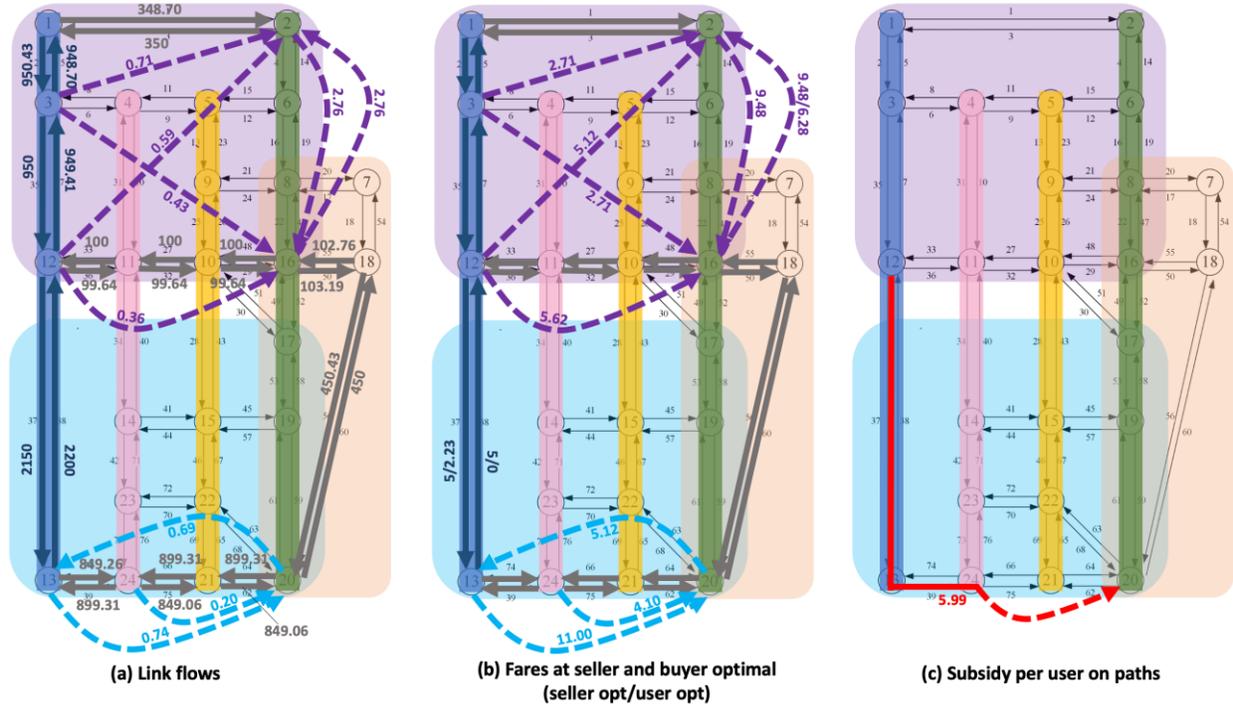

Figure 10. Heterogeneous Demand Case with 5% Optimality Gap Control (links with one fare have same fares at seller and buyer-optimal, links with no fares have 0 fares at both seller and buyer-optimal).

These results that capture endogenous market entry/exit decisions, subsidy decisions, multimodal cooperative behavior, all while respecting user UE path flow principles, elastic demand, individual operator rationality principles, and congestion effects, demonstrates what our model can achieve that cannot be done with any of the generalized Nash equilibrium models in the literature.

# 6 Conclusion

In this study, we propose a MaaS platform design method for MaaS systems which incorporates fixed-route transit services and MOD services, with dummy links that model users' choice of not being a part of the system. The matching problem is formulated as a convex multicommodity network design problem. We derive local stability conditions corresponding to this matching problem and define the optimum that is further constrained to be locally stable to be a globally stable MaaS platform equilibrium. We show that certain variables for this solution to be unique, allowing for market design analysis.

An exact solution algorithm is proposed to solve the matching problem. The algorithm has a branch and bound framework. Each branch is solved through subgradient optimization with a Frank-Wolfe algorithm in each iteration. We prove that due to congestion effects, matchings may be locally unstable. A heuristic is derived from the exact algorithm to solve for a feasible solution to the MaaS platform equilibrium. Local stability is guaranteed by finding a solution that is either naturally stable or subsidized.

Two sets of numerical examples are conducted. With the smaller toy case, we verify the model and algorithm, showing how different locally stable solutions may exist of which one is the globally stable MaaS platform equilibrium. The larger case involves an expanded version of Sioux Falls with 82 nodes and 748 links. We show, among other insights, the following:

- The proposed model can quantify changes in operating costs for one operator in a MaaS platform inducing other operators' market entry decisions due to coopetitive effects.
- The proposed model can quantify impacts of congestion on operators' design decisions.
- A reduction in operating cost to an operator may not have any impact on the market equilibrium due to local stability requirement.
- The modeling of heterogeneous user groups can capture inequities in the market where operators enter primarily to serve higher income users while leaving more lower income users out of the platform.

The method could be applied to service design of MaaS ecosystems, including network design, pricing, and subsidy design. The method is able to incorporate services with different levels of flexibility: fixed-route transit (e.g. subway, bus) and MOD (e.g. ride-hailing, ride-sharing, bikesharing), and ensure the stability of the designed system. We can use this model to design external subsidies, fare bundles, different service region designs in the context of a mobility ecosystem, user heterogeneity, and evaluate the impact of a new external mode (which would alter the dummy link travel disutilities by altering the logsums).

The current bottleneck to the computation cost is the shortest path finding. In the subgradient optimization, different OD pairs have different network costs. Hence, the shortest path finding has to be run as one-to-one. When the number of user groups is large, computation cost can grow significantly if high accuracy is required, requiring modeling trade-offs in zone aggregation. Computation cost can be controlled by adjusting the tolerable gap between the upper and lower bounds in branch and bound, and tolerance parameters of subgradient optimization and Frank-Wolfe algorithm. If there is a tolerable budget of subsidized system total cost, the branch and bound can be terminated when a solution below the budget is found.

In the current modeling framework, we neglected the impact of traffic congestion on MOD by using complete graphs to model MOD service with fixed in-vehicles travel times representing shortest paths between OD pairs. The model could be extended to use a street network to model MOD services; the change would require operators choosing paths to send their vehicular flows that would cover the ODs for passenger paths, and these vehicular path flows would then be bundled along road links to compute congestion effects. Another next step is to model heterogeneous demand as a stochastic many-to-many assignment game. One other possible next step is to define a desired stable outcome space considering policy goals such as efficiency, equity, and so on. In the current model, cost allocation is done in the entire stable outcome space. Subsidy is injected only to stabilize unstable matchings. With a desired stable outcome space specified, platform/agency intervention can be designed accordingly, including subsidy, fare bundles, service restrictions, and so on. The model framework can also be applied to other types of markets: electric vehicle charging infrastructure with multiple shared users, airlines and urban deliveries. For example, urban air mobility markets can be studied by breaking the trip demand into segments by trip purpose and income level to observe where vertiports should arise and which modes they can best work together with or compete against.

## Acknowledgments

This research was conducted with partial support from NSF CMMI-1652735.

## Author statement

The authors confirm contribution to the paper as follows: study conception and design: B. Liu, J.Y.J. Chow; analysis and interpretation of results: B. Liu, J.Y.J. Chow; draft manuscript preparation: B. Liu, J.Y.J. Chow. All authors reviewed the results and approved the final version of the manuscript.

Appendix A. Link parameters of the Sioux Falls network (All costs in the unit of $)

| i | j | $t_{ij}$ | $c_{ij}$ | $w_{ij}$ | i | j | $t_{ij}$ | $c_{ij}$ | $w_{ij}$ | i | j | $t_{ij}$ | $c_{ij}$ | $w_{ij}$ |
|---|---|---|---|---|---|---|---|---|---|---|---|---|---|---|
| 1 | 2 | 6 | 0 | 25900 | 18 | 16 | 3 | 0 | 19680 | 10 | 9 | 3 | 400 | 13916 |
| 2 | 1 | 6 | 0 | 25900 | 18 | 20 | 4 | 0 | 23403 | 10 | 15 | 6 | 400 | 13512 |
| 3 | 4 | 4 | 0 | 17111 | 19 | 15 | 3 | 0 | 14565 | 11 | 4 | 6 | 400 | 4909 |
| 4 | 3 | 4 | 0 | 17111 | 20 | 18 | 4 | 0 | 23403 | 11 | 14 | 4 | 400 | 4877 |
| 4 | 5 | 2 | 0 | 17783 | 20 | 21 | 6 | 0 | 5060 | 12 | 3 | 4 | 400 | 23403 |
| 5 | 4 | 2 | 0 | 17783 | 20 | 22 | 5 | 0 | 5076 | 12 | 13 | 3 | 400 | 25900 |
| 5 | 6 | 4 | 0 | 4948 | 21 | 20 | 6 | 0 | 5060 | 13 | 12 | 3 | 400 | 25900 |
| 6 | 5 | 4 | 0 | 4948 | 21 | 24 | 3 | 0 | 4885 | 14 | 11 | 4 | 400 | 4877 |
| 7 | 8 | 3 | 0 | 7842 | 22 | 20 | 5 | 0 | 5076 | 14 | 23 | 4 | 400 | 4925 |
| 8 | 7 | 3 | 0 | 7842 | 22 | 23 | 4 | 0 | 5000 | 15 | 10 | 6 | 400 | 13512 |
| 8 | 9 | 10 | 0 | 5050 | 23 | 22 | 4 | 0 | 5000 | 15 | 22 | 3 | 400 | 9599 |
| 9 | 8 | 10 | 0 | 5050 | 24 | 13 | 4 | 0 | 5091 | 16 | 8 | 5 | 400 | 5046 |
| 10 | 11 | 5 | 0 | 10000 | 24 | 21 | 3 | 0 | 4885 | 16 | 17 | 2 | 400 | 5230 |
| 10 | 16 | 4 | 0 | 4855 | 1 | 3 | 4 | 400 | 23403 | 17 | 16 | 2 | 400 | 5230 |
| 10 | 17 | 8 | 0 | 4994 | 2 | 6 | 5 | 400 | 4958 | 17 | 19 | 2 | 400 | 4824 |
| 11 | 10 | 5 | 0 | 10000 | 3 | 1 | 4 | 400 | 23403 | 19 | 17 | 2 | 400 | 4824 |
| 11 | 12 | 6 | 0 | 4909 | 3 | 12 | 4 | 400 | 23403 | 19 | 20 | 4 | 400 | 5003 |
| 12 | 11 | 6 | 0 | 4909 | 4 | 11 | 6 | 400 | 4909 | 20 | 19 | 4 | 400 | 5003 |
| 13 | 24 | 4 | 0 | 5091 | 5 | 9 | 5 | 400 | 10000 | 21 | 22 | 2 | 400 | 5230 |
| 14 | 15 | 5 | 0 | 5128 | 6 | 2 | 5 | 400 | 4958 | 22 | 15 | 3 | 400 | 9599 |
| 15 | 14 | 5 | 0 | 5128 | 6 | 8 | 2 | 400 | 4899 | 22 | 21 | 2 | 400 | 5230 |
| 15 | 19 | 3 | 0 | 14565 | 8 | 6 | 2 | 400 | 4899 | 23 | 14 | 4 | 400 | 4925 |
| 16 | 10 | 4 | 0 | 4855 | 8 | 16 | 5 | 400 | 5046 | 23 | 24 | 2 | 400 | 5079 |
| 16 | 18 | 3 | 0 | 19680 | 9 | 5 | 5 | 400 | 10000 | 24 | 23 | 2 | 400 | 5079 |
| 17 | 10 | 8 | 0 | 4994 | 9 | 10 | 3 | 400 | 13916 | | | | | |

Appendix B. OD demand for the Sioux Falls network

| OD ID | Origin | Destination | Demand | OD ID | Origin | Destination | Demand |
|---|---|---|---|---|---|---|---|
| 1 | 2 | 1 | 100 | 16 | 1 | 18 | 100 |
| 2 | 12 | 1 | 200 | 17 | 2 | 18 | 100 |
| 3 | 18 | 1 | 100 | 18 | 12 | 18 | 200 |
| 4 | 13 | 1 | 500 | 19 | 13 | 18 | 100 |
| 5 | 20 | 1 | 300 | 20 | 20 | 18 | 400 |
| 6 | 1 | 2 | 100 | 21 | 1 | 13 | 500 |
| 7 | 12 | 2 | 100 | 22 | 2 | 13 | 300 |
| 8 | 18 | 2 | 100 | 23 | 12 | 13 | 1300 |
| 9 | 13 | 2 | 300 | 24 | 18 | 13 | 100 |
| 10 | 20 | 2 | 100 | 25 | 20 | 13 | 600 |
| 11 | 1 | 12 | 200 | 26 | 1 | 20 | 300 |
| 12 | 2 | 12 | 100 | 27 | 2 | 20 | 100 |
| 13 | 18 | 12 | 200 | 28 | 12 | 20 | 400 |
| 14 | 13 | 12 | 1300 | 29 | 18 | 20 | 400 |
| 15 | 20 | 12 | 500 | 30 | 13 | 20 | 600 |